\documentclass[draft,grl]{AGUTeX_for_PNAS}

\newcommand{\beginsupplement}{%
        \setcounter{table}{0}
        \renewcommand{\thetable}{S\arabic{table}}%
        \setcounter{figure}{0}
        \renewcommand{\thefigure}{S\arabic{figure}}%
     }

\usepackage{lineno}
\usepackage[usenames,dvipsnames]{color}
\usepackage{epstopdf}
\usepackage{amsmath}
\usepackage{graphicx}
\usepackage{natbib}
\usepackage{graphicx}
\setkeys{Gin}{draft=false}
\authorrunninghead{KITE AND RUBIN}

\titlerunninghead{SUSTAINING ENCELADUS' ERUPTIONS}

\begin{document}

%
%

\title{Sustained eruptions on Enceladus explained by turbulent dissipation in tiger stripes}

%
%

%
%



\authors{Edwin S. Kite\altaffilmark{1} \& Allan M. Rubin\altaffilmark{2}} 

\altaffiltext{1}{Department of Geophysical Sciences, University of Chicago, Chicago, IL 60637, USA (kite@uchicago.edu).}
\altaffiltext{2}{Department of Geosciences, Princeton University, Princeton, NJ 08542, USA.}


%
%


\begin{abstract}

{Spacecraft observations suggest that the plumes of Saturn's moon Enceladus draw water from a subsurface ocean, but the sustainability of conduits
linking ocean and surface is not understood. 
Observations show sustained (though tidally modulated) fissure eruptions throughout each orbit, and since the 2005 discovery of the plumes. Peak plume flux lags peak tidal extension by $\sim$1 radian, suggestive of resonance. Here we show that a model of the tiger stripes as tidally-flexed slots that puncture the ice shell can simultaneously explain the persistence of the eruptions through the tidal cycle, the phase lag, and the total power output of the tiger stripe terrain, while suggesting that the eruptions are maintained over geological timescales. The delay associated with flushing and refilling of \emph{O}(1) m-wide slots with ocean water causes erupted flux to lag tidal forcing and helps to buttress slots against closure, while tidally pumped in-slot flow leads to heating and mechanical disruption that  staves off slot freeze-out. Much narrower and much wider slots cannot be sustained. In the presence of long-lived slots, the 10$^6$-yr average power output of the tiger stripes is buffered by a feedback between ice melt-back and subsidence to \emph{O}(10$^{10}$) W, which is similar to the observed power output, suggesting long-term stability. Turbulent dissipation makes testable predictions for the final flybys of Enceladus by the \emph{Cassini} spacecraft. Our model shows how open connections to an ocean can be reconciled with, and sustain, long-lived eruptions. Turbulent dissipation in long-lived slots helps maintain the ocean against freezing, maintains access by future Enceladus missions to ocean materials, and is plausibly the major energy source for tiger stripe activity. [The Proceedings of the National Academies of Sciences version of this paper is available at PNAS Online: \\ http://www.pnas.org/content/113/15/3972.abstract]. }

\end{abstract}

%
%

%

\begin{article}

Enceladus' tiger stripes have been erupting continuously since their discovery in 2005 \citep{Porco2014, Hansen2011, Dong2011, Spitale2015}. The eruptions have been sustained for much longer than that: Saturn's E-ring, which requires year-on-year replenishment from Enceladus, has been stable since its discovery in 1966. Each of the four eruptive fissures is flanked by $<$1 km-wide belts of endogenic thermal emission (10$^4$~W/m for the $\sim$ 500 km total tiger stripe length), a one-to-one correspondence indicating a long-lived internal source of water and energy \citep{NimmoSpencer2013}. The tiger stripe region is tectonically resurfaced, suggesting an underlying mechanism accounting for both volcanism and resurfacing, as on Earth. Enceladus' 30$\pm$10 km thick ice shell is probably underlain by an ocean or sea of liquid water \citep{Thomas2016}, and Enceladus' plume samples a salty liquid water reservoir containing $^{40}$Ar, ammonia, nano-silica, and organics \citep{Postberg2011,Waite2009, Iess2014, Hsu2015}. A continuous connection between the ocean and the surface is the simplest explanation for these observations. However, the consequences of this connection for ice-shell tectonics have been little explored. The water table within a conduit would be $\sim$3.5 km below the surface (from isostasy), with liquid water below the water table, and rapidly ascending vapor plus entrained water droplets above. Condensation of this ascending vapor on the vertical walls of the tiger stripe fissures releases heat that is transported to the surface thermal-emission belts by conduction through the ice shell (Fig. 1) \citep{Porco2014,AbramovSpencer2009,Nimmo2014,Postberg2009}. Because this vapor comes from the water table there is strong evaporitic cooling of water $<~$ 100m below the water table. 
Freezing at the water table could release latent heat but would clog the fissures with ice in $<$1 yr. This energy deficit has driven consideration of shear-heating, intermittent eruptions, thermal-convective exchange with the ocean, and heat-engine hypotheses \citep{NimmoSpencer2013, Matson2012, Nimmo2007}. It is easiest to explain the observations if the heat is made within the plumbing system.

\begin{figure}
\centering

\noindent\includegraphics[width=0.9\columnwidth,trim=5mm 20mm 5mm 2mm, clip=true]{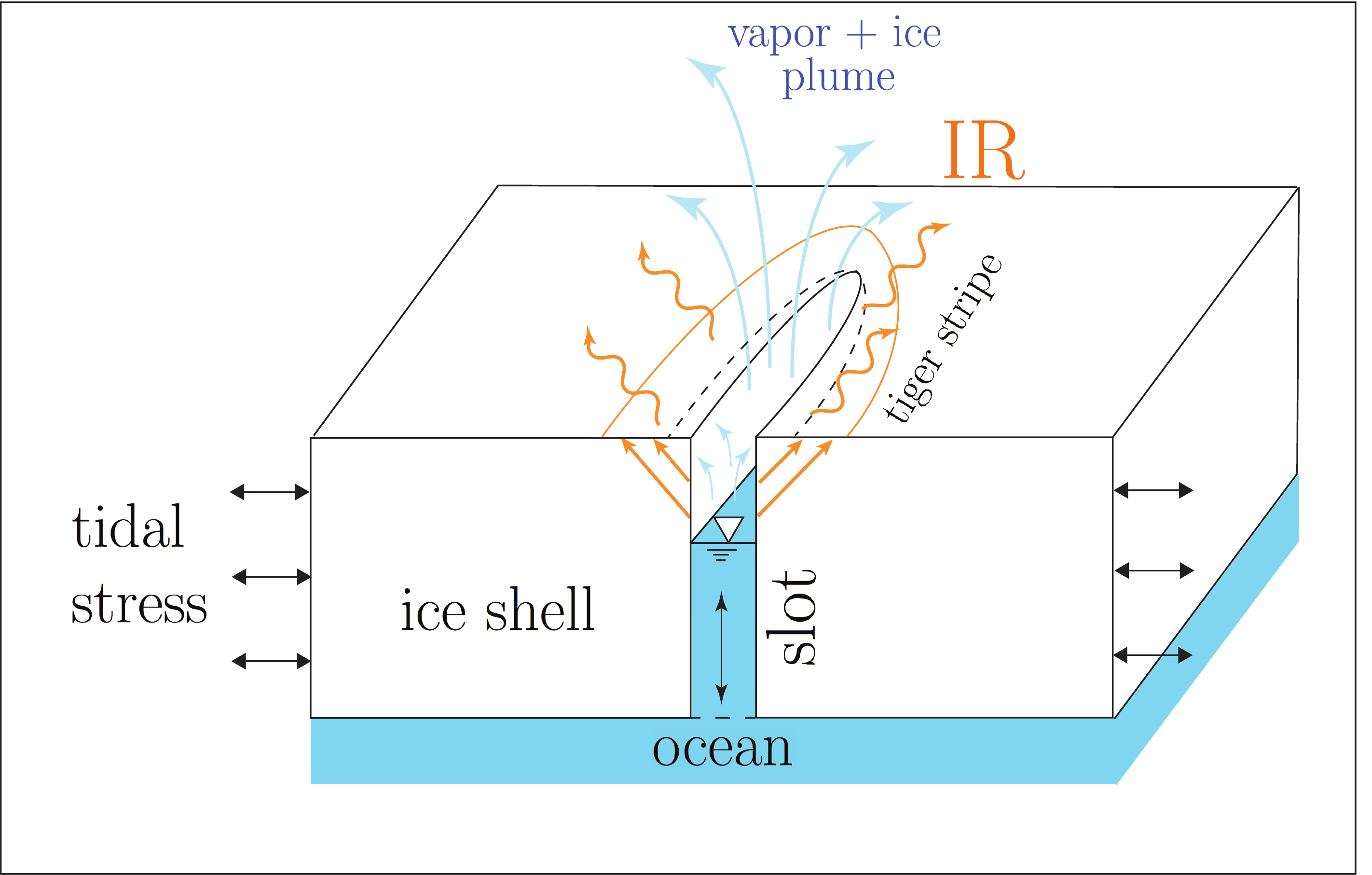}
\caption{The erupted flux from Enceladus (blue arrows) varies on diurnal timescales, which we attribute to daily flexing (dashed lines) of the source fissures by Saturn tidal stresses (horizontal arrows). Such flexing would also drive vertical flow in slots underneath the source fissures (vertical black arrow), which through viscous dissipation generates heat. This heat helps to maintain the slots against freeze-out despite strong evaporitic cooling by vapor escaping from the water table (downward-pointing triangle). The vapor  ultimately provides heat (via condensation) for the envelope of warm surface material bracketing the tiger stripes (orange arrows; ``IR'' corresponds to infrared cooling from this warm material).}
\label{figure_label}
\end{figure}

\begin{figure}
\centering
\includegraphics[width=1.0\columnwidth,trim=5mm 0mm 0mm 0mm, clip=true]{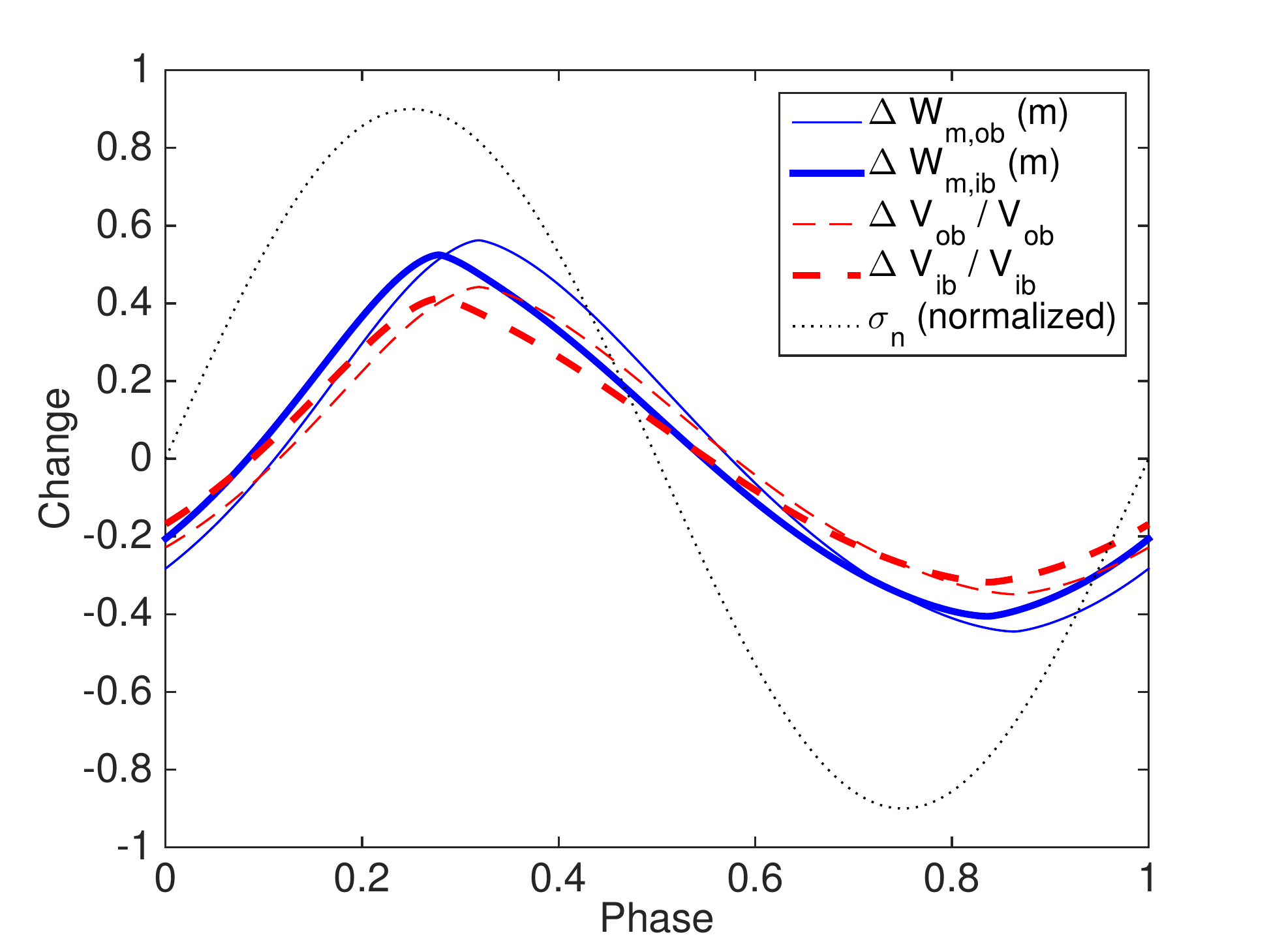}
\caption{Tidal flexing cycle for interacting slots assuming 2 inboard ($ib$) and 2 outboard ($ob$) slots, $E$ = 6 GPa, $L$ = 100 km. Slot half-width $W_0$ = 1 m. $\Delta W_m$ is maximum width change, $\Delta V/V$ is fractional change in slot water volume, and $\sigma_n$ is extensional stress to 90\% of its own peak amplitude. 
\label{diurnalcycle}}
\end{figure}

\begin{figure}
\vspace{0.05in}
\centering
\includegraphics[width=0.7\columnwidth,trim=20mm 10mm 0mm 10mm]{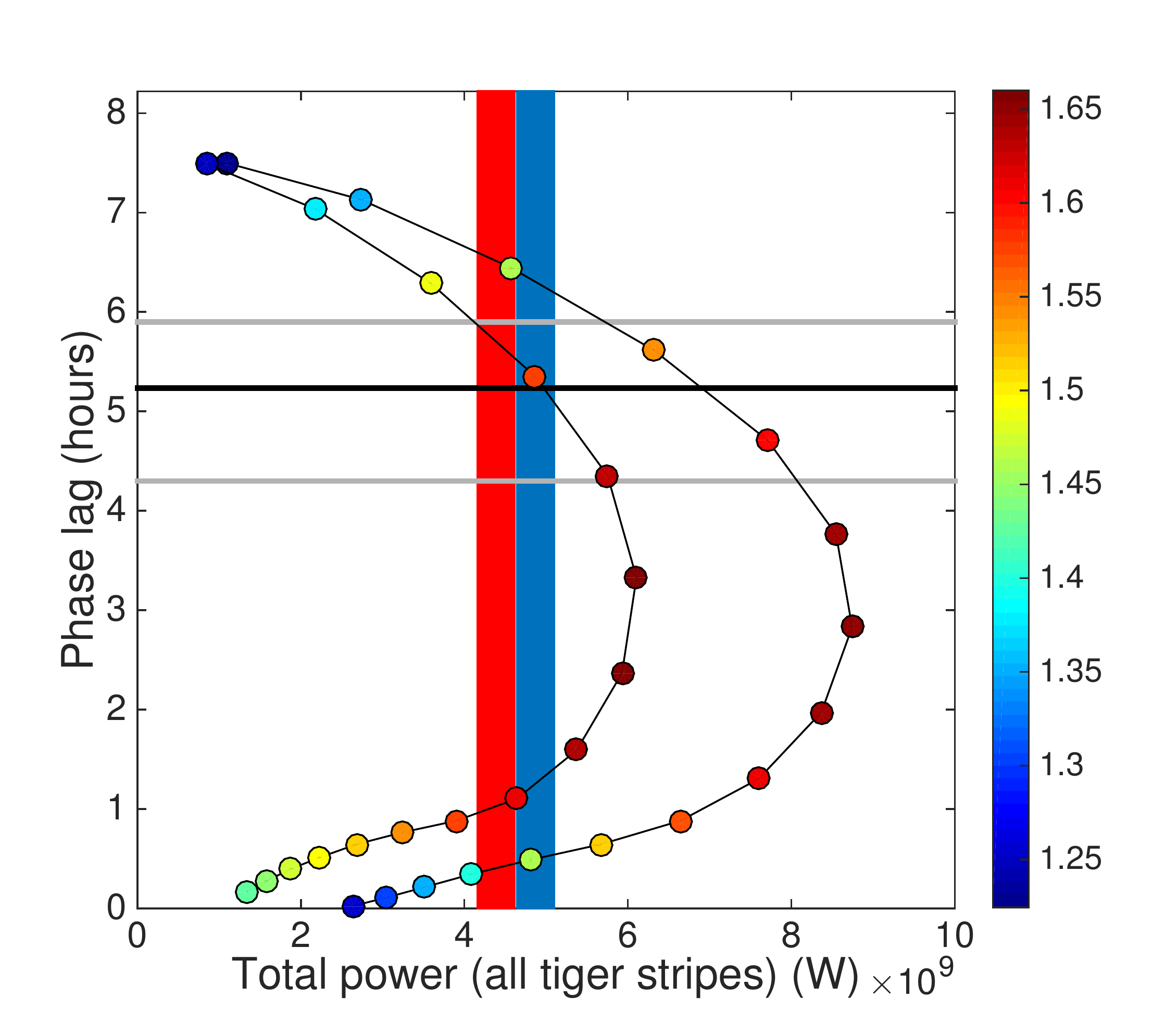}
\caption{Turbulent dissipation that can account for the observed power output of Enceladus (red bar; blue bar includes the additional power inferred for reheating of cold ice at depth) can also match the observed phase lag of Enceladus (thick black horizontal line shows phase lag relative to a fiducial model, gray lines show 1$\sigma$ error in observed phase lag).  Curves correspond to time-averaged power for (left curve) four slots of equal length 100~km, and (right curve) 93~km length outboard slots and 151~km length inboard slots.  Dot color corresponds to the fractional change in aperture (max/min) during the tidal cycle. For each curve, initial (tidal-stress-free) half-width is sampled at 0.125 m (uppermost dots) and then at 0.125 m intervals up to 2~m (lowermost dots). Half-widths that best fit both the phase delay and power constraints are 0.375-0.625 m. \label{powervsphaselag}}
\end{figure}

The observed long-term steadiness of ice and gas output is modulated (for ice) by fivefold variability at the period $p$~=~33~hours of Enceladus' eccentric orbit about Saturn (the diurnal tidal period). Peak eruptive output anomalously lags peak tidal extension (by 5.1$\pm$0.8 hours relative to a fiducial model of the tidal response), and fissure eruptions continue from all four tiger stripes at Enceladus' periapse  when all tidal crack models predict that eruptions should cease \citep{Porco2014,Spitale2015,Nimmo2014,Hurford2009}. The sustainability of water eruptions on Enceladus affects the moon's habitability (e.g., \citep{McKay2014}), as well as astrobiology (follow-up missions to Enceladus could be stymied if the plumes shut down). Despite the importance of understanding the sustainability of the eruptions, basic questions remain open: How can eruptions continue throughout the tidal cycle? How can the liquid water conduits obtain the energy to stay open -- as needed to sustain eruptions -- despite evaporitic cooling and viscous ice inflow? Why is the total power of the system $\sim$5~GW~(not~$\sim$0.5~GW~or~$\sim$50~GW)? Do tiger stripe mass and energy fluxes drive ice shell tectonics, or are the tiger stripes a passive tracer of tectonics? 

We have found that a simple model of the fissures as open conduits can simultaneously explain both the maintenance of Enceladus' eruptions throughout the tidal cycle and the sustainability of eruptions on 10$^{-1}$--10$^1$ yr timescales, while predicting that eruptions are sustained over 10$^6$ yr timescales. Fissures are modeled as parallel rectangular slots with length $L$ $\approx$ 130 km, depth $Z$ = 35 km, stress-free half-width $W_0$, and spacing $S$ = 35 km. Slots are connected to vacuum at the top, and open to an ocean at the bottom (Fig. 1, SI Appendix). Subject to extensional slot-normal tidal stress $\sigma_n$ = (5$\pm$2) $\times$ 10$^4$ $\mathrm{sin}(2 \pi t / p)$ Pa modified by elastic interactions between slots, the water table initially falls, water is drawn into the slots from the ocean (which is modeled as a constant-pressure bath), and the slots widen (Fig. \ref{diurnalcycle}). Wider slots allow stronger eruptions because the flow is supersonic and choked \citep{Schmidt2008}. Later in the tidal cycle, the water table rises, water is flushed from the slots to the ocean, the slots narrow, and eruptions diminish (but never cease) (Fig. \ref{diurnalcycle}). 
Solving the coupled equations for elastic deformation of the icy shell with turbulent flow of water within the tiger stripes allow us to compute $W(t)$ (SI Appendix). $W_0$~$>$~2.5~m slots oscillate in phase with $\sigma_n$, $W_0$~$<$~0.5 m slots lag $\sigma_n$ by $\sim$$\pi/2$ rad, and resonant slots ($W_0$~$\sim$~1~m, tidal quality factor $\sim$ 1) lag $\sigma_n$ by $\sim$ 1~radian (Fig. S4). The net liquid flow feeding the eruptions ($<$10 $\mu$m/s) is much smaller than the peak tidally-pumped vertical flow ($\sim$ $\pm$1~m/s for $W_0$~$\sim$ 1~m, $Re$~$>$~10$^5$). Although the amplitude of the cycle in water table height is reduced when the slot is hydrologically connected to the ocean relative to a hypothetical situation where the slot is isolated from the ocean, the flow velocity, driven by the deviation of the water table from its equilibrium elevation, is very much larger than in the hydrologically isolated case.

Turbulent liquid water flow into and out of the slots generates heat. Water temperature is homogenized by turbulent mixing, allowing turbulent dissipation to balance water table losses and prevent icing-over.  Ice forming at the water table is disrupted by aperture variations and vertical pumping; water cooled by evaporation, if sufficiently saline, will sink and be replaced by warmer water from below. A long-lived slot must satisfy the heat demands of evaporitic cooling at the water table (about 1.1$\times$ the observed IR emission; Methods) plus heating and melt-back of ice driven into the slot by the pressure gradient between the ice and the water in the slot \citep{CuffeyPatterson2010} (SI Appendix). Turbulent dissipation can balance this demand for $W_0$~=~(1$\pm$0.5) m, corresponding to phase lags of 0.5-1 rad, consistent with observations. Eruptions are then strongly tidally-variable but sustained over the tidal cycle, also matching observations. $W_0$ $<$ 0.5 m slots freeze shut, and $W_0$ $>$ 2.5 m slots would narrow. Power output is sensitive to the amplitude $k$ of conduit roughness, which is poorly constrained for within-ice conduits. For the calculations in this paper we use $k$ = 0.01 m; for discussion see SI Appendix. Near-surface apertures $\sim$10 m wide are suggested by modeling of high-temperature emission \citep{Goguen2013}, consistent with near-surface vent flaring \citep{Mitchell2005}. Rectification by choke points \citep{Schmidt2008} (which are required to explain the absence of sodium in the gas plume; \citep{Schneider2009}), together with condensation on slot walls, and ballistic fall-back \citep{Postberg2011} could plausibly amplify the $<$2-fold slot-width variations in our model to the 5-fold variations in the flux of ice escaping Enceladus.\footnote{For outflow velocities of 300-500 m/s, near-surface vent temperatures of $\sim$200K \citep{Goguen2013}, and rapid vent wall / gas pressure equilibration \citep{IngersollPankine2010}, the effective mean fracture width implied by UV occultation constraints on vapor flux (200 kg/s) is only $\sim$1 mm for 5 $\times$ 10$^5$m total fracture length. The aperture of the surface fissures presumably widens and narrows along strike.}  Water's low viscosity slows the feedback that causes the fissure-to-pipe transitions for silicate eruptions on Earth  \citep{BruceHuppert1989,Wylie1999}, which is suppressed for Enceladus by along-slot mixing (SI Appendix).

The mass and heat fluxes associated with long-lived slots would drive regional tectonics (SI Appendix). Slow inflow of ice into the slot \citep{Rothlisberger1972} occurs predominantly near the base of the shell, where ice is warm and soft. Inflowing ice causes necking of the slot, which locally intensifies dissipation until inflow is balanced by melt-back. Melt-back losses near the base of the shell cause colder ice from higher in the ice shell to subside. Because subsidence is fast relative to conductive warming timescales, subsidence of cold more-viscous ice is a negative feedback on the inflow rate. This negative feedback adjusts the flux of ice consumed by melt-back near the base of the shell to balance the flux of subsiding ice (Fig. S6), which in turn is equal to the mass added by condensation of ice from the vapor phase above the water table (Fig. S6). The steady-state flux of ice removed from the upper ice shell via subsidence and remelting at depth depends on $Z$, $S$, moon gravity, and the material properties of ice. Using an approximate model of ice-shell thermal structure, this steady-state flux is approximately proportional to $Z$ in the range 20 km $<$ $Z$ $<$ 60 km (Fig. S8) and is $\sim$3 ton/s (7 mm/yr subsidence, $Pe$~$\approx$~6) for $Z$ = 35 km. This long-term value is comparable to the inferred post-2005 rate of ice \emph{addition} to the upper ice shell,  2 ton/s (assuming the observed 4.4~$\pm$~0.2~GW cooling of the surface is balanced by re-condensation of water vapor on the walls of the tiger stripes above the water table; \citep{IngersollPankine2010,Nimmo2014,Howett2014, NimmoSpencer2013}). If near-surface condensates are distributed evenly across the surface of the tiger stripe terrain (either by near-surface tectonics \citep{BarrPreuss2010}, or by ballistic fallback), then the balance is self-regulating because increased (decreased) tiger stripe activity will reduce (increase) the rate at which accommodation space for condensates is made available via subsidence in the near surface. This is consistent with sustained eruptions on Enceladus at the \emph{Cassini}-era level over $>$10$^6$ yr. Under these conditions the ice is relatively cold and nondissipative. In summary, turbulent dissipation of diurnal tidal flows (Fig. 1) explains the phase lag and the diurnal-to-decadal sustainability of liquid-water-containing tiger stripes (Fig. \ref{powervsphaselag}), and the coupling between tiger stripes and the ice shell forces a $>$10$^6$ yr geologic cycle that buffers Enceladus' power to approximately the \emph{Cassini}-era value.

Our model makes predictions for the results of \emph{Cassini}'s final flybys of Enceladus. We predict that endogenic thermal emission will be absent between tiger stripes, in contrast to the regionwide thermal emission that is expected if the phase lag is caused not by water flow in slots, but instead by a~$Q$~$\sim$~1~ice~shell \citep{Nimmo2014,Behounkova2015}. In our view, the tiger stripes are the loci of sustained emission because other fractures are too short ($L$ $<$ 100 km) for sustained flow. Because sloshing homogenizes water temperatures along stripe strike (SI Appendix), the magnitude of emission should be relatively insensitive to local tiger-stripe orientation, a prediction that distinguishes the slot model from all crack models. Variations in thermal emission on 10 km length scales have been reported (e.g. \citep{Porco2014}), and might allow this prediction to be tested. The slot model predicts a smooth distribution of thermal emission at $>$km scale. 
Our model is more easily reconciled with curtain eruptions \citep{Spitale2015} than jet eruptions \citep{Porco2014}, and it can provide a physical underpinning for curtain eruptions. Localized emission might still occur, for example near Y-junctions. The pattern of spatial variability in orbit-averaged activity should be steady, in contrast to bursty hypotheses, and vapor flux should covary with ice-grain flux. 
Spatially-resolved variability with orbital phase \citep{Porco2014,Spitale2015} should correspond to the effects of water transfer along-slot, elastic interactions between slots and along slot walls, and possible along-slot width variations. Our basic model might be used as a starting point for more sophisticated models of Enceladus coupling fluid and gas dynamics \citep{IngersollPankine2010}, as well as the tectonic evolution and initiation of the tiger stripe terrain (e.g., \citep{Behounkova2012}). Such coupling may be necessary to understand the initiation of ocean-to-surface conduits on ice moons including Enceladus and Europa, which remains hard to explain \citep{CrawfordStevenson1988,Roth2014}. Initiation may be related to ice-shell disruption during a past epoch of high orbital eccentricity: such disruption could have created partially-water-filled conduits with a wide variety of apertures, and evaporative losses caused by tiger stripe activity would ensure that only the most dissipative conduits (those with $W_0$~=~1$\pm$0.5~m; Fig. \ref{powervsphaselag}) endure to the present day. Eccentricity variations on  $>$10$^7$~yr timescales may also be required if the ocean is to be sustained for the $>$10$^7$~yr timescales that are key to ocean habitability \citep{MeyerWisdom2007,Tyler2011}. Ocean longevity could be affected by heat exchange with self-sustained slots in the ice shell. Testing habitability on Enceladus (or Europa) ultimately requires access to ocean materials, and this is easier if (as our model predicts) turbulent dissipation keeps the tiger stripes open for $\gg$Kyr.

\begin{acknowledgments}
We thank O. Bialik, C. Chyba, W. Degruyter,  S. Ewald, E. Gaidos, T. Hurford, L. Karlstrom, D. MacAyeal, M. Manga, I. Matsuyama, K. Mitchell, A. Rhoden, M. Rudolph, B. Schmidt, K. Soderland, J. Spitale, R. Tyler, and S. Vance; their contributions, ideas, discussion, and suggestions made this work a pleasure. We thank M. B{\u e}hounkov{\'a}, A. Ingersoll and M. Nakajima for sharing unpublished work; and two anonymous reviewers for thorough, timely, and constructive comments. We thank the Enceladus Focus Group meeting organizers. This work was enabled by a fellowship from Princeton University, and by the U.S. taxpayer through grant NNX11AF51G.
\end{acknowledgments}

%
%

\newpage
\appendix 

\beginsupplement

\section{Supplementary Materials.}

%
%
%
%

\subsection{Tidal stress cycle.}

\noindent Enceladus is in 1:1 spin:orbit resonance with Saturn, but its eccentric orbit produces a time-varying (diurnal) stress cycle. We calculate $\sigma_n(t,\phi,\theta)$ for Enceladus' tiger stripes (where $\sigma_n$ is extensional slot-normal tidal stress, $t$ is time, $\phi$ is colatitude and $\theta$ is longitude) using a thin-shell approximation in which radial stresses, physical-libration stresses, obliquity stresses \citep{Hurford2009}, nonsynchronous-rotation stresses, and tiger-stripe interactions are neglected \citep{Nimmo2007, SmithKonterPappalardo2008, Wahr2009}. We set the Poisson ratio to 0.33 and use fiducial Love numbers $l_2$ = 0.04 and $h_2$ = 0.2 for ease of comparison with previous work (e.g., \cite{Nimmo2007}). High Love numbers are appropriate for a thin ice shell above a global ocean \citep{McKinnon2015}. (Section A7 discusses sensitivity to $l_2$ and $h_2$). Tiger stripe locations are traced from \texttt{http://photojournal.jpl.nasa.gov/catalog/PIA12783}. The results (Fig. S1) show that stresses along the tiger stripes are generally ``in phase'' (Fig. S2), with peak compression near periapse. Therefore we approximate the tiger stripes as straight, parallel, and with in-phase forcing. The peak-to-trough amplitude of the stress cycle along the main tiger stripes is (1.31 $\pm$ 0.28) $\times$ 10$^5$ Pa. Sensitivity tests introducing $<$ 90$^\circ$ phase lags between stripes show no significant effect on time-averaged power. 

\begin{figure}
\centering
\includegraphics[width=0.7\textwidth, clip=true, trim=0mm 0mm 20mm 0mm]{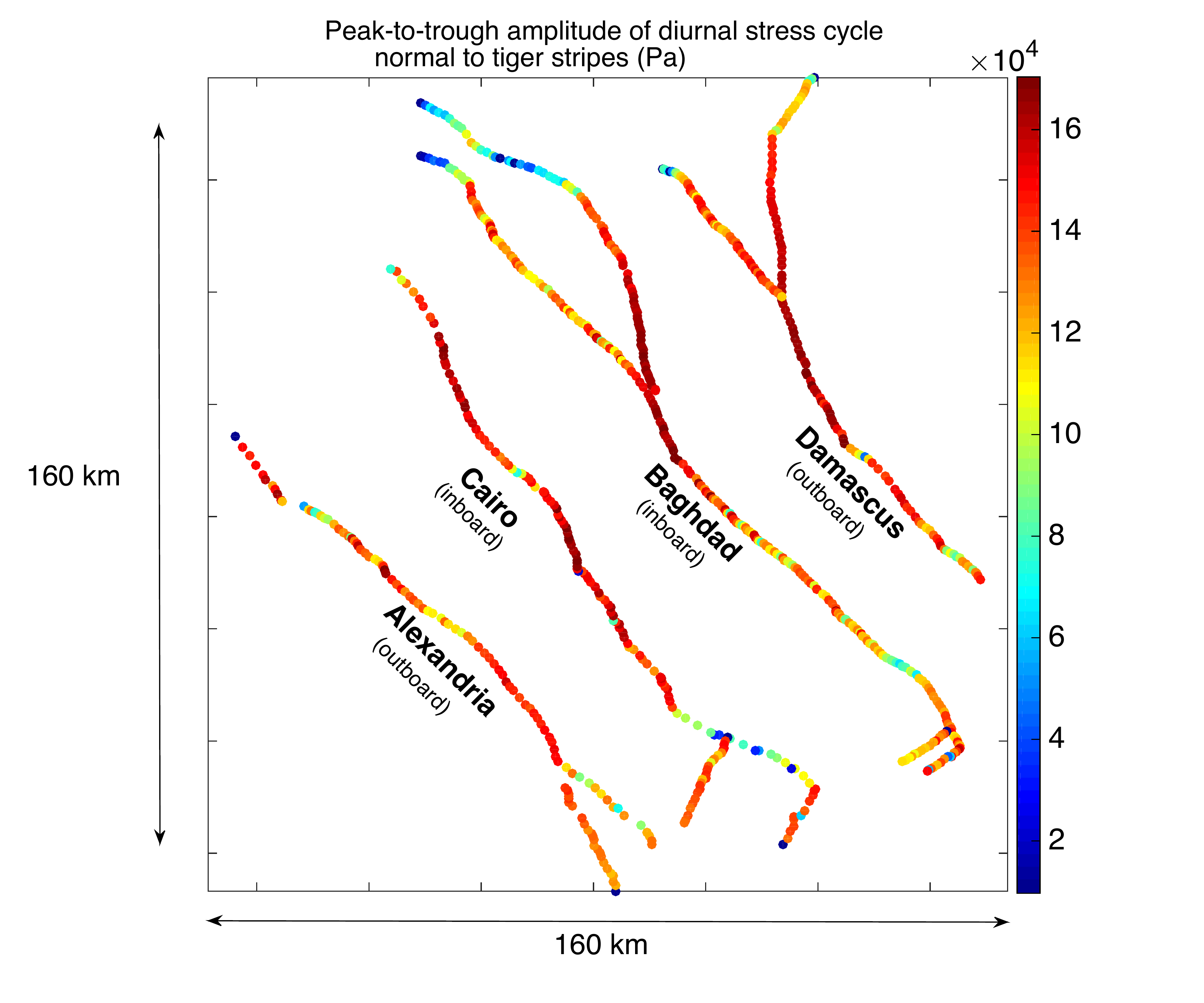}
\caption{Map of the peak-to-trough amplitude of the diurnal cycle in normal stress across Enceladus' tiger stripes for $h_2$ = 0.2, $l_2$ = 0.04. Along-fissure lengths for tiger stripes range (in our mapping) from 133 km (Alexandria \emph{and} Damascus) to 194 km (Cairo), mean 158 km. Great-circle distances between the tips of currently active regions as reported by \cite{Porco2014} range from  79 km (Alexandria) to 165 km (Baghdad), mean 122 km. The tiger stripes shown here are the ones that appear freshest and most prominent in imagery of the surface. There is generally good correlation between topographically-prominent tiger stripes and eruption sources, although \cite{Spitale2015} show that eruptions also emanate from a branch of Baghdad Sulcus that is not particularly geologically prominent (and so is missing from our map).}
\end{figure}

\begin{figure}
\centering
\includegraphics[width=1.0\textwidth]{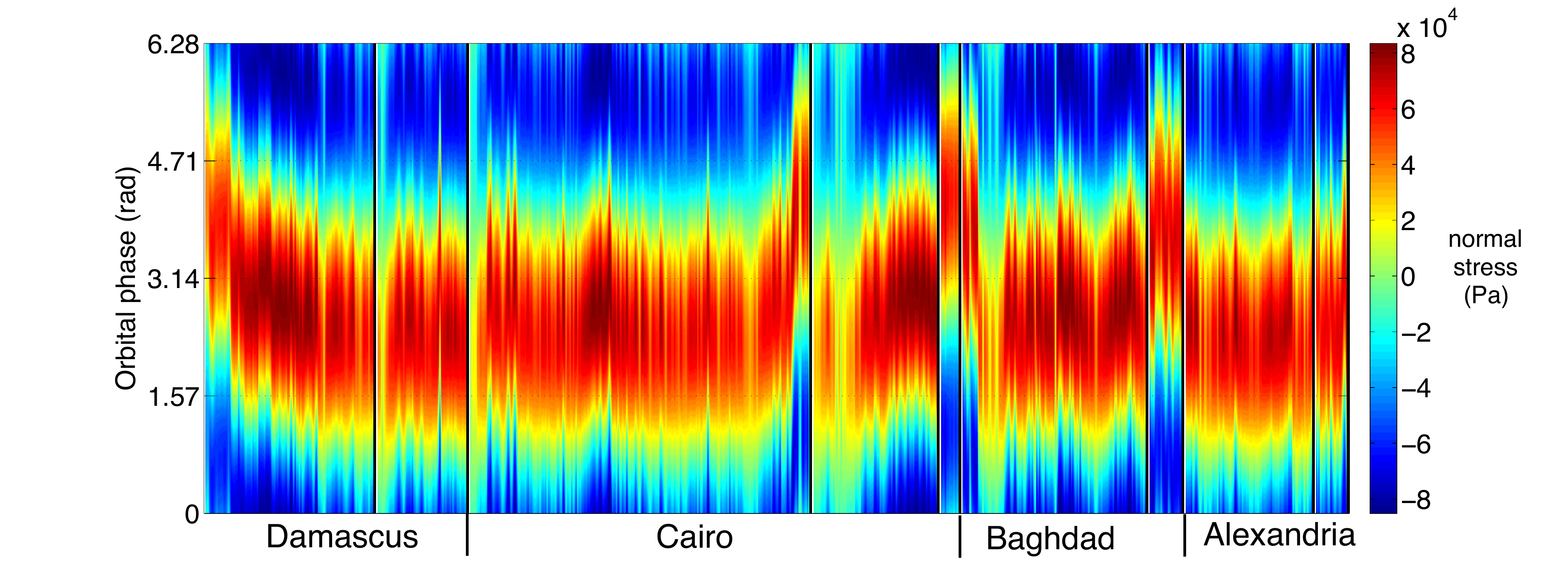}
\caption{Contour plot of tiger stripe normal stresses as a function of phase and relative distance along the tiger stripes for $h_2$ = 0.2, $l_2$ = 0.04. Each vertical line corresponds to a single vertex picked on the tiger stripes.}
\end{figure}

\subsection{Slot width cycle.}

\begin{figure}
\centering
\includegraphics[width=1.0\textwidth, trim={10mm 14mm 4mm 4mm},clip]{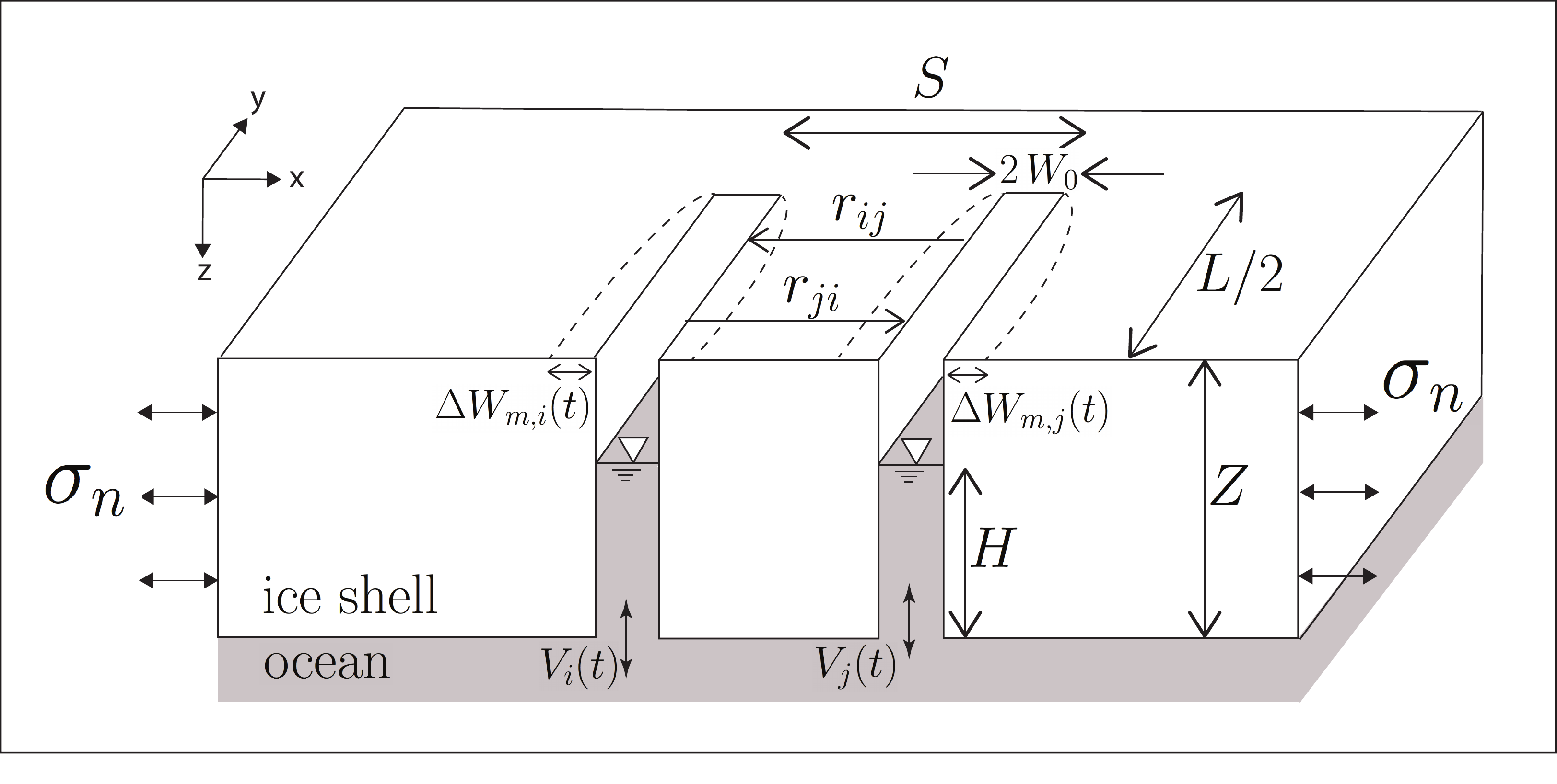}
\caption{Definitions of terms used in model description (\S A.2 - \S A.3). Two parallel rectangular slots of uniform stress-free half-width $W_0$, length $L$, and spacing $S$, subject to tidal stress $\sigma_n(t)$, interact elastically (coefficients $r_{ij}$ and $r_{ji}$). The combination of tidal stress and slot-slot interactions leads to cycles in maximum width change $\Delta W_{m}(t)$. Cycles in water table height are small relative to zero-stress water table height $H$. $\Delta W_{m}(t)$ is coupled to water ingestion/flushing at velocity $V(t)$.}
\end{figure}

\noindent Consider a thin shell of impermeable, isotropic ice I with thickness $Z$ floating on a very voluminous water ocean. Stresses do not vary with depth within the shell, membrane stresses are neglected, and the ocean is treated as a constant-pressure bath. The shell is perforated by one or more rectangular slots of length $L$ and (initially) uniform half-width $W_0$ ($L \gg W_0$), with walls of roughness $k$ (Fig. S3). For the slot calculations we neglect curvature (adopting Cartesian geometry) and assume plane stress. Plane stress is more reasonable for Enceladus' tiger stripes than plane strain: $L$~=~90-190 km (from imaging), whereas $Z$ is most likely 30-40 km and very probably $\leq$~60 km \citep{Iess2014}.
 
Water rises within the slot to $D$ $\approx$ $\frac{\rho_i}{\rho_w} Z$, where $\rho_i$ = 916 kg/m$^3$ is ice density and $\rho_w$~=~1000 kg/m$^3$ is water density. Ocean over-pressure \citep{Wang2006} or siphoning of the water by bubbles could raise the mean water table, but this would have only a small effect on our diurnal-cycle results. (Liquid water is not seen at the surface, suggesting that the ocean is not currently over-pressured by more than 0.3 MPa.) The slot is assumed to have adjusted prior to $t = 0$ through freezing and melt-back so that at $t = 0$ the slot width is uniform with depth. 
  
The ice slab is subject to periodic, extensional, slot-normal stress $\sigma_n$ = $\frac{A}{2}$sin($n_f t$) where $A$ is the peak-to-trough amplitude of the diurnal tidal stress cycle, $n_f$~=~2$\pi/p$, and $p$~$\sim$~33~hours is the period of Enceladus' orbit. Sinusoidal time dependence is valid for small orbital eccentricity. 

In the absence of water, and assuming linear elasticity, the plan-view slot area for a single slot ($Y$) oscillates in phase with the forcing:

\begin{equation} 
Y(t) = 2 W_0 L + \frac{ \pi}{2} L \Delta W_{m} = L \left( 2 W_0 + \frac{ \pi \sigma_n L}{2 E }\right) 
\end{equation}

\noindent where $E$ is the Young's modulus of ice ($\sim$ 6 $\times$ 10$^9$ Pa \citep{Nimmo2004,NimmoManga2009}), and $\Delta W_{m}$ is the change in half-width at the location of greatest (\underline{m}aximum) half-width change. This is because $\Delta W(y_*)$ = 2 $\sigma_n E^{-1} \sqrt{(L/2)^2 - y_*^2}$  where $y_*$ is distance along-slot measured from the middle of the slot length \citep{Gudmundsson2011}.

Adding viscous fluid modifies the tidal cycle in $Y$. Conservation of water volume during a single tidal cycle relates changes in $\Delta W$ and volume exchange with the ocean:

\begin{equation}
\frac{\partial \Delta W_{m}}{\partial t}   \left(\frac{\pi L}{2}\right) \left( D + h \right) + 
\frac{\partial h}{\partial t} \left( 2 W_0 L + \frac{\Delta W_{m} \pi L }{2} \right)  = 
Q
\end{equation}

\noindent where $h$ is the time-varying part of the water level and $Q$ corresponds to influx at the slot base. We assume $h \neq h(y)$, which is reasonable for Enceladus (\S A.2, \S A.6, \S A.10), and we neglect terms of order $h/z$ or $(z-D)/z$.
The first term on the left-hand side accounts for width change, and the second term accounts for changes in $h$. Slot-wall freeze-on and melt-back has characteristic timescale $\gg$ $p$ (\S A.6). $h$ is set by the time-varying normal stress that would act across the crack in the absence of relative motion of the crack walls ($\sigma_\infty$) as modified by the crack-wall stress perturbation due to relative motion ($\sigma_{el}$): 

\begin{equation}
\rho_w g h = -\sigma_\infty - \sigma_{el} = - \frac{A}{2} \mathrm{sin}\left( n_f t \right) + \Delta W_{m}(t) \frac{2  E}{L} 
\end{equation}

\noindent where the pressure distribution within the water column responds to changes in far-field stresses or in water volume on timescale $\ll$ $p$ (inertia is unimportant).\footnote{This can be seen by considering an inviscid fluid column sealed at the bottom and subject instantaneously to a water-pressure gradient of $P_{max} / H(t=0)$ $\approx$ $ P_{max}$ / 0.9$Z$. The column adjusts to $H(t = \tau_{adj}) = P_{max} / (\rho_w g)$ where $\tau_{adj}$ is an adjustment timescale.  Euler's equation of inviscid motion gives $H(t) \sim (P_{max} / \rho_w H) t^2$, so $\tau_{adj} \sim \sqrt{H/g}$.  Viscosity does not greatly alter these conclusions for the solutions that match $Cassini$ data (Fig. 2). The water table tracks the rising pressure when $\tau_{adj} / p \ll 1$. For $Z$ $\sim$ 20 km and $P_{max}$ = 1 bar, this ratio of timescales is \emph{O}(10$^{-3}$) for Enceladus and \emph{O}(10$^{-4}$) for Europa; negligibly small. }  \mbox{Differentiating,}

\begin{equation}
\rho_w g \frac{\partial h}{\partial t} = - n_f \frac{A}{2} \mathrm{cos}\left( n_f t \right) + \frac{\partial \Delta W_{m} (t)}{\partial t}\frac{2 E}{L}
\end{equation}


\noindent Substituting,

\begin{multline}
\left(D+\frac{1}{\rho_w g} \left(-\frac{A}{2}\mathrm{sin}(n_f t) + \frac{\Delta W_{m} E}{L/2}\right)\right)\frac{\partial \Delta W_{m}}{\partial t} \left(\frac{\pi L}{2}\right) 
+ \\ 
\frac{1}{\rho_w g} \left(-n_f \frac{A}{2} \mathrm{cos}(n_f t) + \frac{\partial \Delta W_{m}}{\partial t} \frac{2E}{L}\right)\left(2 W_0 L + \frac{\pi L \Delta W_{m}}{2}\right) = Q
\end{multline}

\noindent Rearranging,

\begin{equation}
\frac{\partial \Delta W_{m}}{\partial t} = 
\frac{4 g \rho_w Q + 4 n_f L W_{0}  A \mathrm{cos}(n_f t)
+ \pi n_f L \Delta W_{m}  A  \mathrm{cos}(n_f t)  }
{16EW_{0} + 8\pi E\Delta W_{m} - \pi L A \mathrm{sin}(n_f t) + 2 \pi g  \rho_w D L} 
\end{equation}


\noindent As a check we run $Q$ = 0, $W_0$ = 1 m. This corresponds to a slot that contains water, but is hydrologically isolated from the ocean. We find $\Delta W_{m}$~$\sim$~0.02 m, as expected, which would lead to negligible tidal flow within the slot and negligible viscous heat generation. However, hydrological connections to the ocean are suggested by the observation of ocean material in the plume \citep{Postberg2011,Waite2009,Hsu2015}.

Solving (A6) requires requires $Q(t)$ = $Y(t) V(t)$ ($V$ is velocity), but first we consider interactions between slots. 

\subsection{Slot-slot interactions.}
\noindent The tiger stripes are $S$ $\sim$ 35 km apart and $L$ $\sim$ 130 km long (stripes can be mapped using images of the surface or alternatively using the trace of individually triangulated supersonic ice sources; $L$ is slightly shorter for the second approach), so elastic interactions between slots can be important (Fig. S1). We model interactions using a two-dimensional displacement discontinuity implementation of the boundary element method \citep{CrouchStarfield1983, RubinPollard1988}, with plane stress, neglecting planetary curvature. Using this code, we investigated interactions between slots of a variety of geometries, branching patterns, and lengths. The tiger stripes are reasonably well-approximated as being straight, parallel, equally spaced (Fig. S1) and in phase (Fig. S2). With those approximations, the tiger stripe terrain is essentially symmetric, so that the deformation cycle for the Alexandria and  Damascus Sulci (the outboard slots) will be the same, and the deformation cycle for the Baghdad and Cairo Sulci (the inboard slots) will be the same. However, the deformation cycle for the outboard slots will differ from that for the inboard slots. Because slot width is much less than tiger-stripe spacing, the coupling terms are insensitive to both possible between-slot variations in mean width and possible along-slot variations in width.

We can write the stress change that would occur on the walls of the perturbed stripe-pair if those walls were not free to move as follows:

\begin{equation}
\Delta P_{ij} = \frac{\Delta w_i}{\left( \frac{\partial w_i}{\partial p_i} \right)}  \left( \frac{\partial w_j}{\partial p_i} \right) \left( \frac{\partial p_j}{\partial w_j} \right)
 = \left( \frac{\partial p_{j,iso}}{\partial w_{j,iso}} \right) r_{ij} \Delta W_{m,i}
\end{equation} 

\noindent where the subscript $i$ corresponds to the perturbing stripe-pair, the subscript $j$ corresponds to the perturbed stripe-pair, and the subscript $iso$ corresponds to deformation of a \emph{single} slot of length equal to $j$. ($\partial p_{iso} / \partial w_{iso} = 2 E / L $ for a straight slot.) We average over slot plan-view shape variations by defining $w$ as the along-slot average width of the slot. (This averaging anticipates the along-slot averaging in our discharge-velocity calculation, \S A.4). The second term in brackets ($\partial w_j / \partial p_i$) evaluates the thickness change of $j$ due to a pressure change at $i$ for the case of constant stress within $j$, i.e. allowing the walls of $j$ to undergo relative motion. $\Delta w_i$ is found iteratively, and all other terms are pre-computed using the two-dimensional displacement-discontinuity boundary element code. Elastic interactions are rapid relative to the tidal cycle (the sound-crossing timescale $(3 S / v_{sound}) = (10^2\,\,\mathrm{km} / 4\,\,\mathrm{km\, s^{-1}})~\approx~30\,\, \mathrm{s} \ll p$). 

To take account of interacting slots, (A3)-(A6) must be modified. For one of a pair of cracks ($i$) that interacts with another pair ($j$), $\sigma_{\infty,i}$ must be modified to include the normal stress perturbation on $i$ due to relative motions of the walls of $j$ (enforcing zero relative motion of the walls of $i$), while $\sigma_{el,i}$ must be modified to account for the other of the $i$ pair. We obtain (by analogy with A3-A6)

\begin{equation*}
\sigma_{\infty,i} = \frac{A}{2} \mathrm{sin}(n_f t) + \frac{2E}{L}r_{ji}\Delta W_{m,j} 
\end{equation*}
\begin{equation*}
\sigma_{el,i} = \frac{2E}{L}r_{ii}\Delta W_{m,i} 
\end{equation*}
\begin{equation*}
\rho_w g h_i = -\frac{A}{2}  \mathrm{sin}(n_f t) + \frac{2E}{L}(r_{ii} \Delta W_{m,i} - r_{ji}\Delta W_{m,j} )
\end{equation*}
\begin{multline*}
\left(D+\frac{1}{\rho_w g} \left(-\frac{A}{2}\mathrm{sin}(n_f t) + \frac{2E}{L}\left(r_{ii} \Delta W_{m,i} - r_{ji} \Delta W_{m,j} \right)\right)\right)\frac{\partial \Delta W_{m,i}}{\partial t} \left(\frac{\pi L}{2}\right) 
+ \\ 
\frac{1}{\rho_w g} \left(-n_f \frac{A}{2} \mathrm{cos}(n_f t) +  \frac{2E}{L}\left( r_{ii} \frac{\partial \Delta W_{m,i}}{\partial t} - r_{ji} \frac{\partial \Delta W_{m,j}}{\partial t} \right)\right)\left(2 W_0 L + \frac{\pi L \Delta W_{m,i}}{2}\right) = Q_i
\end{multline*}

and rearranging yields an implicit equation analogous to (A6):

\begin{equation}
\frac{\partial \Delta W_{m,i}}{\partial t}  = \frac{Q_i + \left(\rho_w g\right)^{-1} \left(\frac{1}{2}A n_f \mathrm{cos}(n_f t) + \frac{2}{L} E r_{ji} \frac{\partial \Delta W_{m,j}}{\partial t} \right) Y }{\frac{1}{2} L \pi \left(D - \left(\rho_w g\right)^{-1}\left( \frac{1}{2} A \, \mathrm{sin}(n_f t) - \frac{2E}{L}(r_{ii} \Delta W_{m,i}  - r_{ji} \Delta W_{m,j} ) \right) \right) + \frac{2 E r_{ii} Y}{L g \rho_w}}
\end{equation}

\noindent where $r_{ii}$ corresponds to the interaction between paired tiger stripes, and $r_{ji}$ corresponds to the interaction between the outboard slot pair and the inboard slot pair (we use $r_{ji}$ to refer to the influence of the motion of slot $i$ on $j$). The $r$ terms are computed for uniform stress changes within the slots, consistent with the the assumption of in-phase tidal loading. When the deformation is in phase between paired tiger stripes, $r_{ii}$ is in the range 0 $<$ $r_{ii}$ $<$ 1 (deformation-promoting) and $r_{ji}$ is negative (deformation-retarding).

For the solutions that match data (Fig. 2), the slots track each other within 20\% of power output and within 1 hr in phase. However, the slot-pairs can deform very differently for parameters that may be encountered during Enceladus' geologic evolution (\S A.7). 

Interactions between slots may be relevant to the origin of the parallel, equally-spaced tiger stripes: once a slot is established, it will produce bending stresses in the adjacent shell that favor failure at horizontal distances approximately equal to the shell thickness (e.g., \cite{Porco2014}). This preferred wavelength is found for icebergs on Earth \citep{Reeh1968}.

\subsection{Turbulent viscous dissipation.}
\noindent Extensional (compressional) stress causes $h$ to fall (rise) which draws water into the slot from the ocean (expels water from the slot into the ocean). Velocity $V$ at the slot inlet depends on the pressure gradient $\nabla P$, the roughness of the slot walls $k$, and the Reynolds number $Re$. Here we linearize by assuming $\nabla P~\sim~(D~+~h)^{-1} (\sigma_n(t)~-~2~(r_{ii} \Delta W_{m,i}~-~r_{ji} \Delta W_{m,j}) E / L)$. Velocities normal to the crack wall are much smaller than velocities parallel to the crack wall, and we also ignore flow parallel to the long axis of the tiger stripes ; in detail the pressure gradient will be affected by (and feed back on) slot-width variations with depth in the shell. Combining the Darcy-Weisbach and Colebrook-White equations gives (\cite{Nalluri2009}, their Equation 4.10):

\begin{equation}
\left | V \right | = -2 \sqrt{2 g_{E} d S_f} \,\, \mathrm{log_{10}}  \left( \frac{k}{3.7 d} + \frac{2.51 \nu}{d \sqrt{2 g_{E} d S_f}} \right)
\end{equation}

\noindent where $g_{E}$ is Earth gravity, 
$S_f$ is the dimensionless energy loss gradient ($S_f$ = $\nabla P_z / \rho g_{E}$), $\nu$~$\approx$~1.8~$\times$~10$^{-6}$~m$^2$~s$^{-1}$ is the kinematic viscosity of water at 0$^\circ$ C, and we set the equivalent pipe diameter $d$ to be 4 times the mean slot half-width $W_{1/2}^*(t) = \frac{1}{2}\left(2 W_0 + \frac{\pi}{2} \Delta W_{m}\right)$. The first term inside the brackets accounts for rough turbulent flow, which dominates for $\left| V \right| \sim$~1~m/s, $k$~$>$~0.001~m (Enceladus-relevant conditions), so

\begin{equation}
|Q| \approx  - 4 L W_{1/2}^* 
 \sqrt{2 g_{E} ( 4 W_{1/2}^*) 
\left| \frac{\Delta P}
{\left( D - \Delta P \frac{1}{\rho_w g} \right)}  
\frac{1}{\rho_w g_E}\right| }
\,\mathrm{log_{10}} \left( \frac{k}{14.8 W_{1/2}^* } \right)
\end{equation}

\noindent where 

\begin{equation}
\Delta P =\left( \frac{A}{2} \right)  \mathrm{sin}(n_f t) - (r_{ii} \Delta W_{m,i} - r_{ji} \Delta W_{m,j}) 2E / L
\end{equation}

\noindent and we have again linearized the pressure gradient. The sign of $Q$ (positive for inflow into the slot) is set equal to the sign of $\nabla P$.


We now have an expression for $Q$ which can be substituted in (Eqn. A8) to solve for $\Delta W_{m}(t)$. By multiplying $Q$ by the magnitude of the pressure difference between the ocean and the slot, we obtain an estimate of $F(t)$, the average power dissipated per length of tiger stripe (see \S A.10 for a more detailed discussion). 
An alternative approach (from \cite{Tritton1988}, their  Fig. 2.11) gives the nondimensional average pressure gradient $\xi$ as the maximum of $\xi_l$ (laminar) and $\xi_t$ (turbulent) 

\begin{equation*}
\xi = \mathrm{max}(\xi_{l}, \xi_{t}) 
\end{equation*}
\begin{equation*}
\xi_{l} = 32 Re 
\end{equation*}
\begin{equation}
\xi_{t} = 0.25 Re^{\sim1.75} 
\end{equation}

\noindent where $Re$ = 2$W^*_{1/2}(t)$ $V(t)$ / $\nu$. 
The power per unit length (W/m) is then (\cite{Tritton1988}, their Fig. 2.11)

\begin{equation}
F \approx |V| \frac{D \xi \mu^2}{\rho_w (2 W_0 + (\pi/2)\Delta W_{m} )^2}
\end{equation}

\noindent where we neglect the difference between $D$ and $D + h$, assume a static pressure gradient, and advect water parcels through the gradient at speed $|V|$. $\mu$ is the dynamic viscosity of water ($\mu$~=~$\nu$~$\rho_w$). We find that (A9-A10) gives the same power output as (A13) for $k$ $\approx$ 0.02 m, $W_0$ = 1 m. 

Most of this power will go (at steady state) into evaporative losses at the water table or into the ice, rather than being mixed into the ocean: heat flow scales as 1/$\delta_{bl}$, and boundary layer thicknesses $\delta_{bl}$ are small within the slot. In addition, heated water is buoyant relative to cold water of the same salinity (for salinity $>$ 2 \%, which is plausible for Enceladus waters that have undergone fractional evaporation) so it will remain under the slots (either as a sheet, or concentrated in cupolas beneath slots).  The total power for the slot is $L F(t)$, and because the change in water temperature during one tidal cycle is $\ll$ 1 K, the diurnal-mean $F$ is used to compare to observations (Figs. 2-3).

Fig. S4 shows results for uncoupled slots, $W_0$~=~$\{ 0.25, 0.375, 0.5, ... 2.5\}$~m, $L$~=~100 km, $S$~=~$Z$ = 35 km.

\begin{figure*}
\centering

a)
\includegraphics[width=0.45\textwidth,trim=0mm 0mm 5mm 10mm, clip=true]{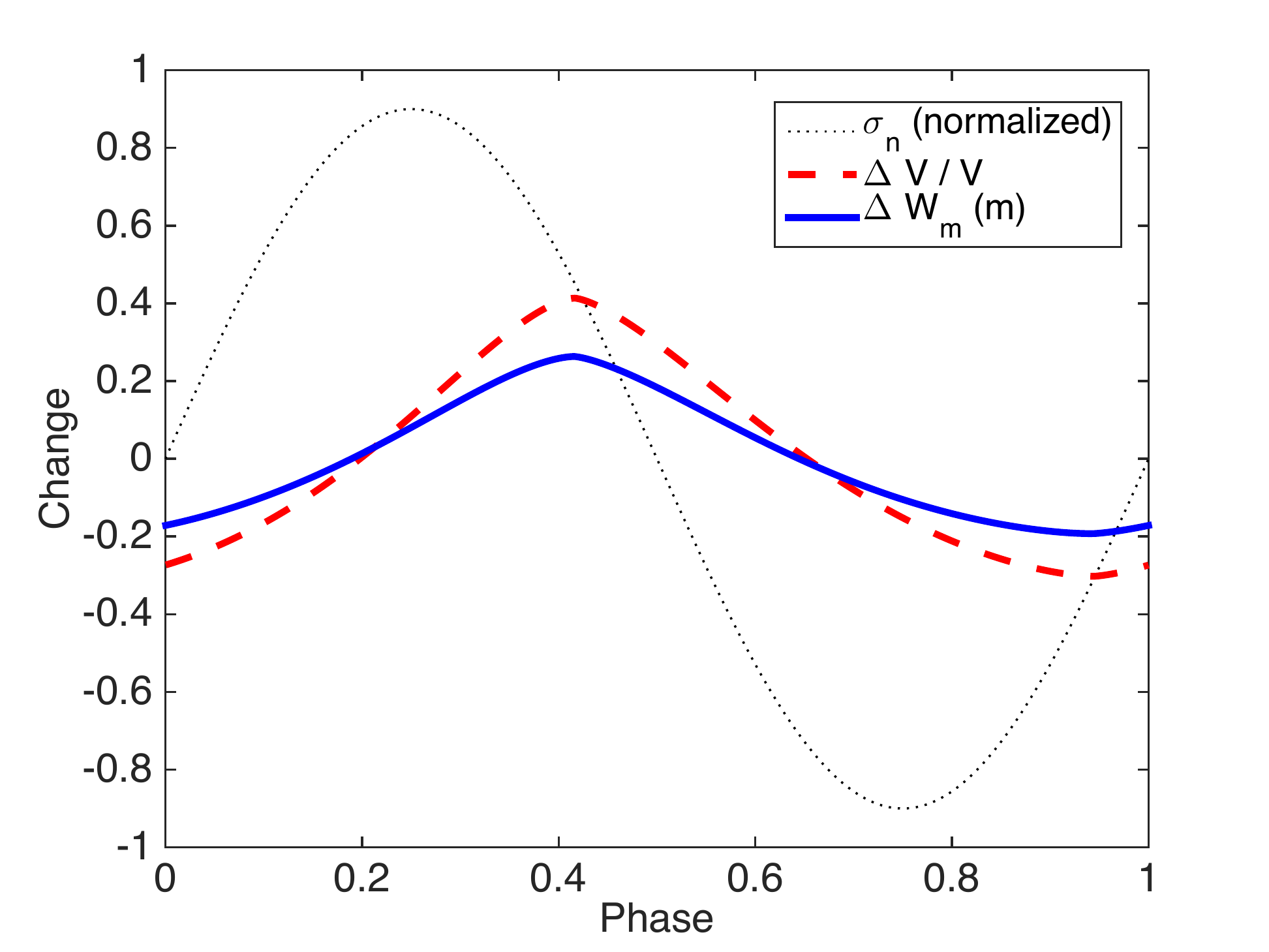}

b)
\includegraphics[width=0.45\textwidth,trim=0mm 0mm 5mm 10mm, clip=true]{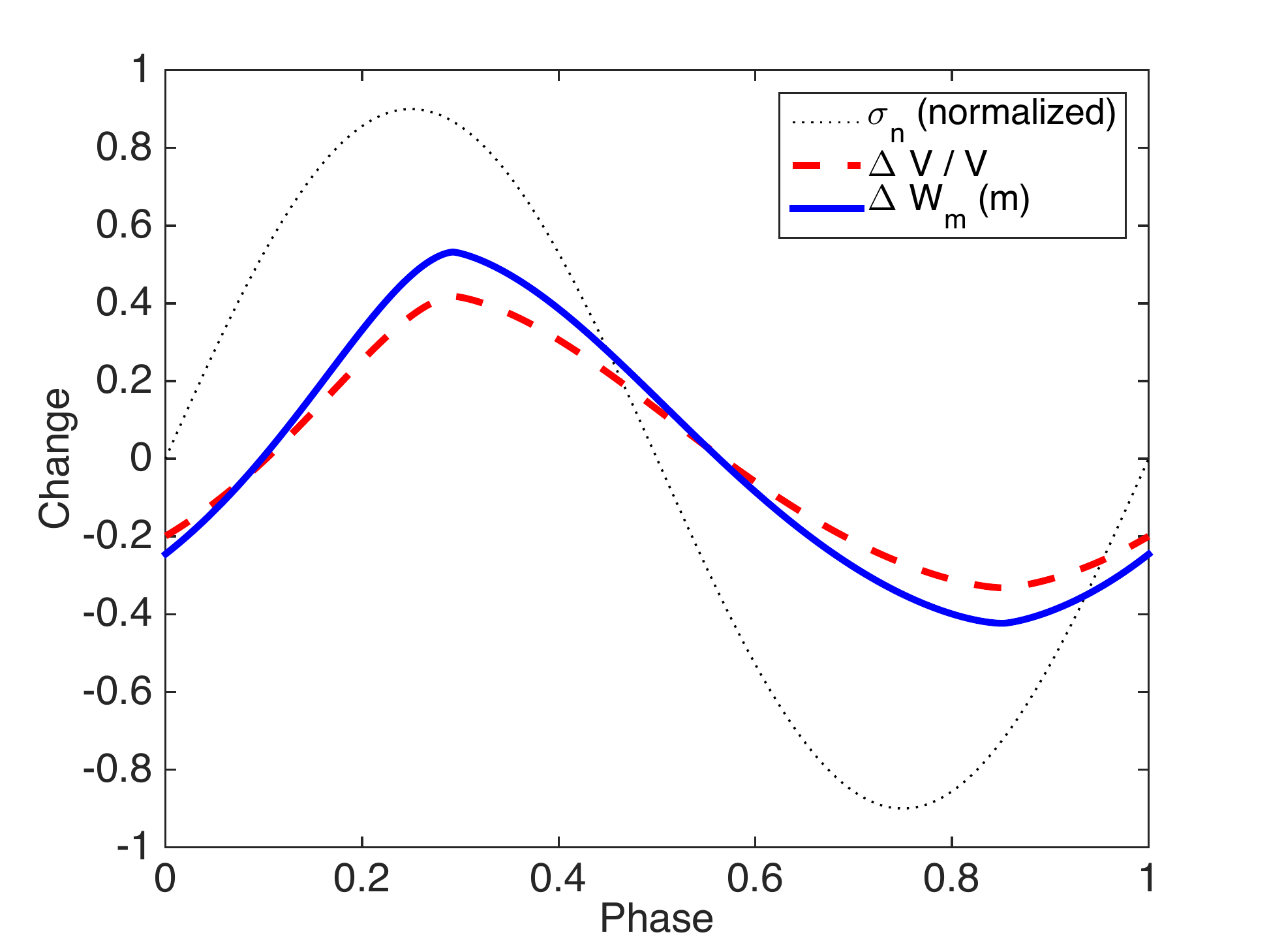}

c)
\includegraphics[width=0.45\textwidth,trim=0mm 0mm 5mm 10mm, clip=true]{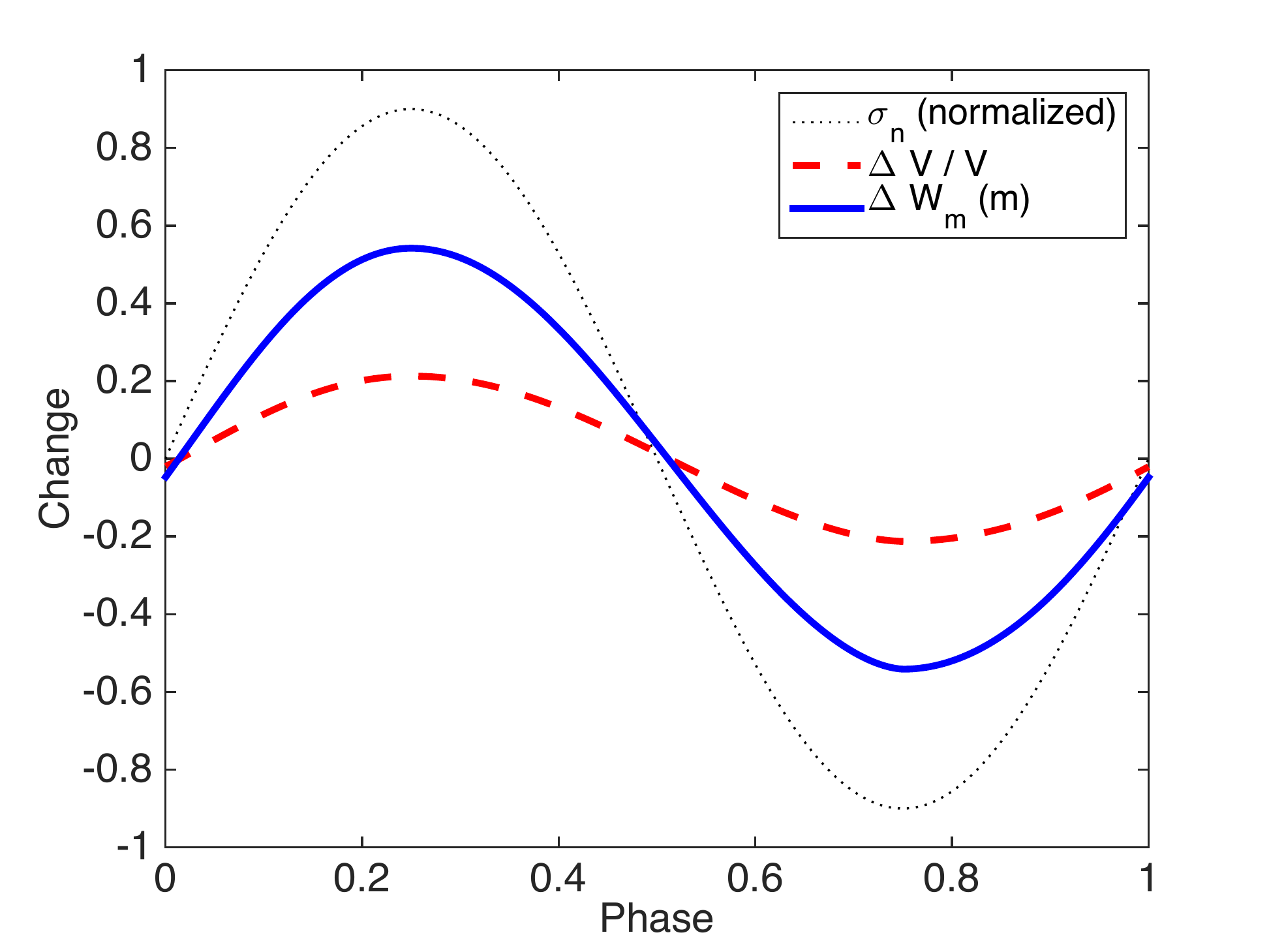}

\end{figure*}


\begin{figure}
\centering

d)
\includegraphics[width=0.5\textwidth]{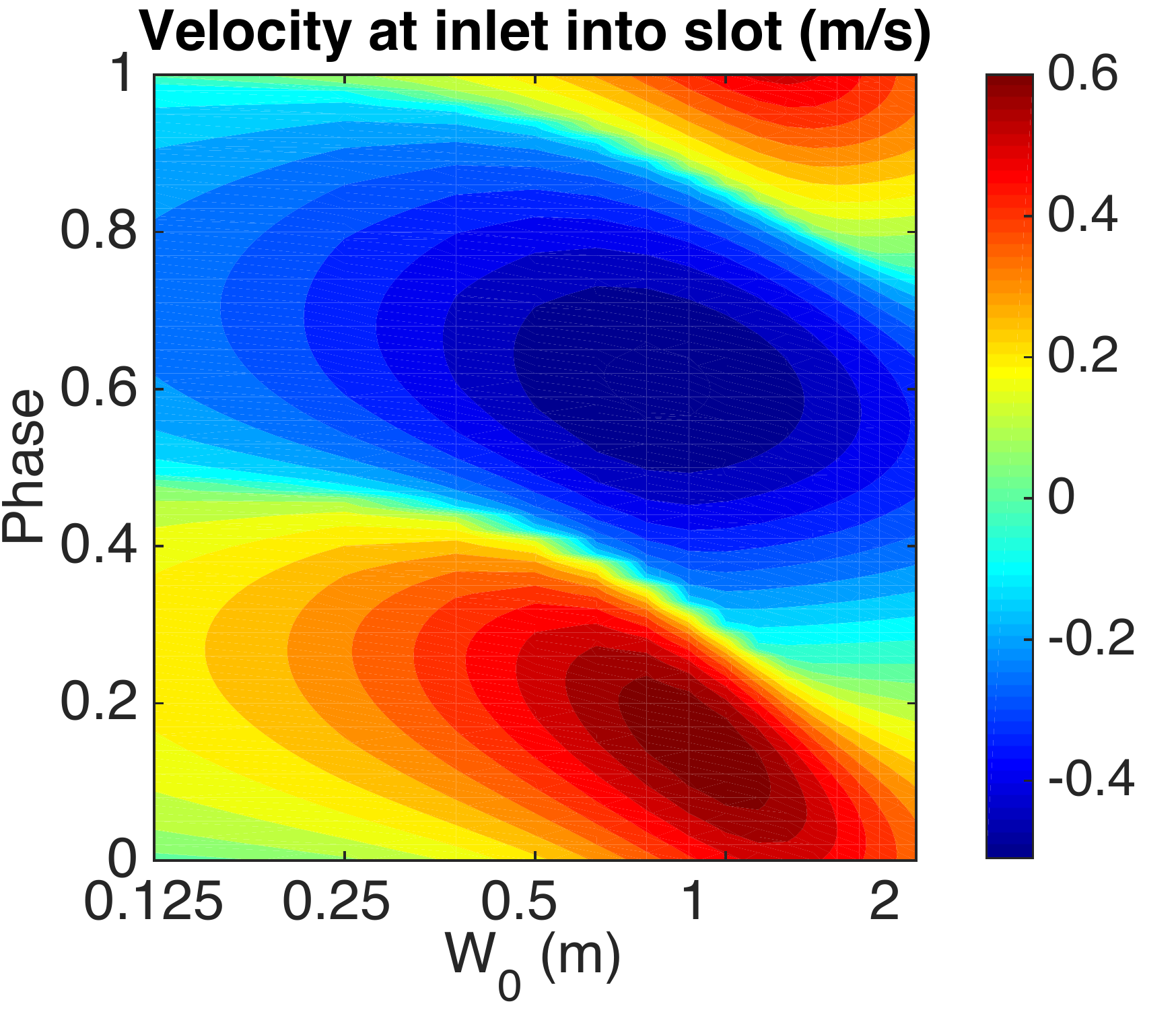}

e)
\includegraphics[width=0.7\textwidth]{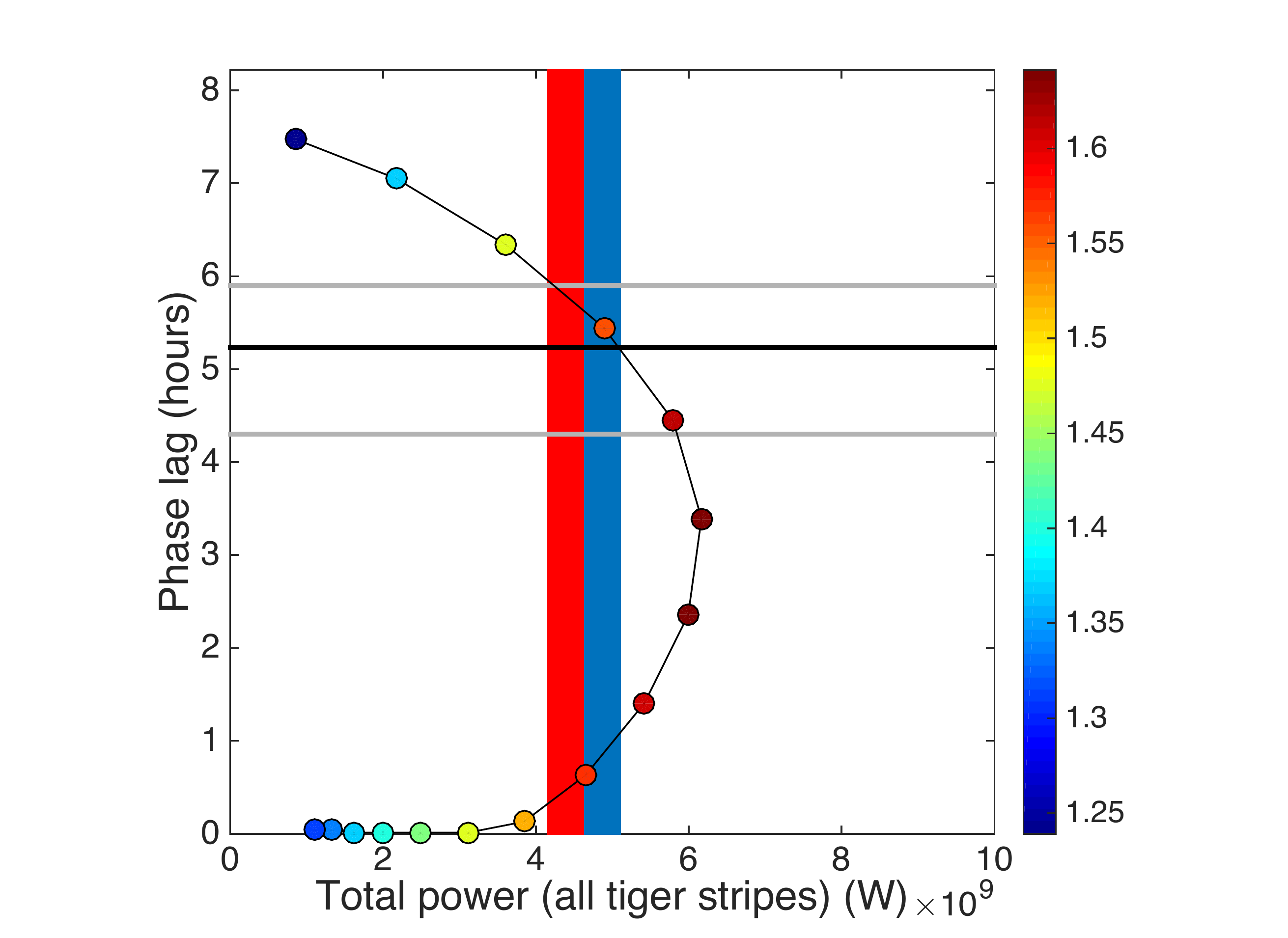}
\caption{ \small
Tidal flexing cycle for slots that do not interact elastically (for 4 slots each with $L$ = 100 km, $E$ = 6 GPa). \textbf{(a)} Slot half-width $W_0$ = 0.5 m. $\Delta W_m$ is maximum width change, $\Delta V/V$ is fractional change in slot water volume, and $\sigma_n$ is extensional stress normalized to 90\% of its own peak amplitude. \textbf{(b)} As (a), but $W_0$ = 1.0 m. \textbf{(c)} As (a), but $W_0$ = 2.0 m. \textbf{(d)} Diurnal cycle of velocity into slot at ocean inlet, as a function of initial width. Contour interval 0.05 m/s. \textbf{(e)} Phase lag versus total power output. Initial (tidal-stress-free) half-width is sampled at 0.25 m (uppermost dots) and then at 0.125 m intervals up to 2.5~m (lowermost dots). Dot color corresponds to the fractional change in aperture (max/min) during the tidal cycle. Thick black horizontal line shows phase lag relative to a fiducial model, gray lines show 1$\sigma$ error in observed phase lag. Red bar shows the observed power output of Enceladus; blue bar includes the additional power inferred for reheating of cold ice at depth.}
\end{figure}

Fig. S5 shows results for coupled slots of equal length, $W_0$ = $\{ 0.25, 0.375, 0.5, ... 2.5\}$ m, $L$ = 100 km, $S$ = $Z$ = 35 km.

\begin{figure*}
\centering
a)
\includegraphics[width=0.55\textwidth,trim=0mm 0mm 5mm 10mm, clip=true]{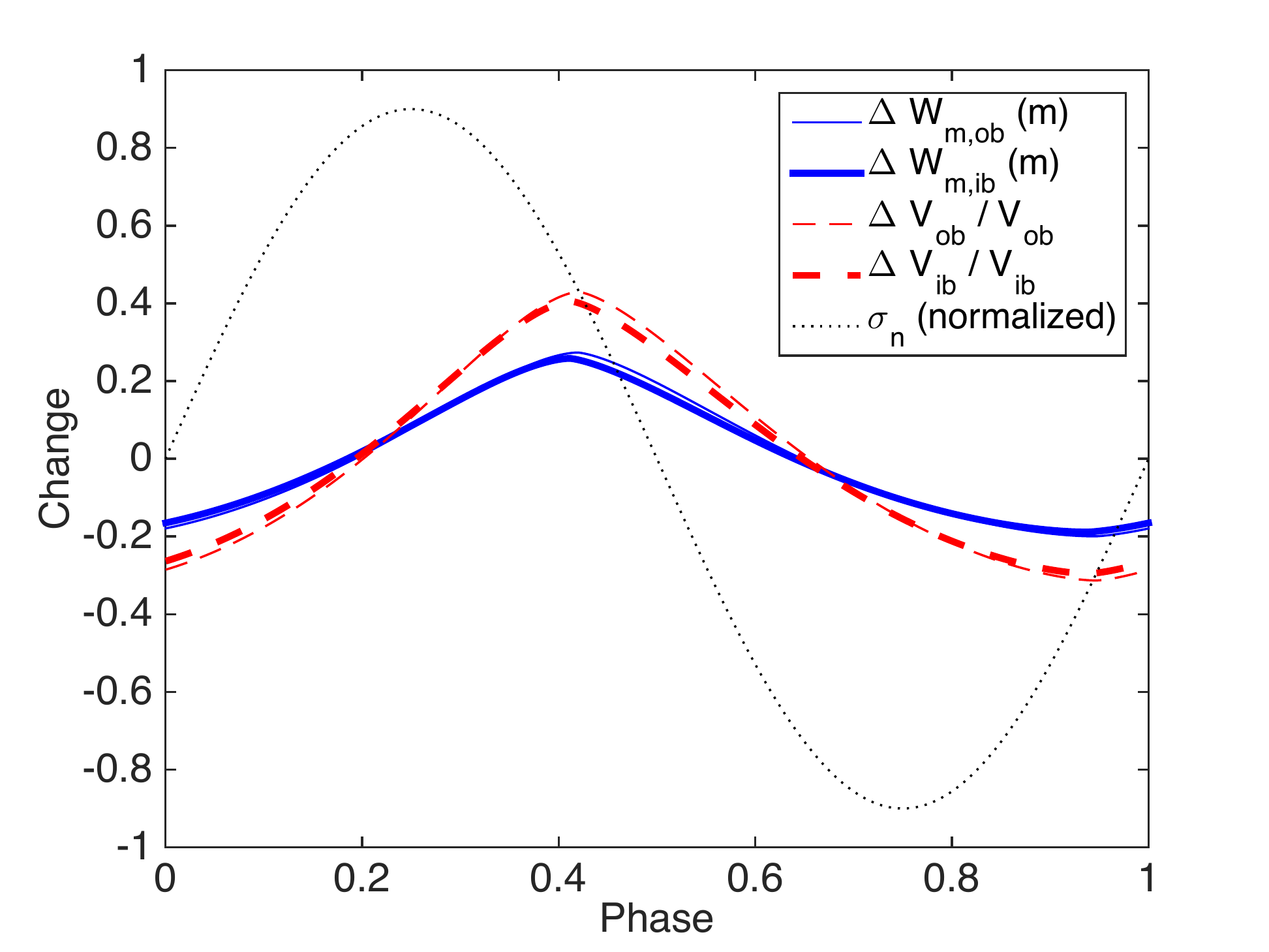}

b)
\includegraphics[width=0.55\textwidth,trim=0mm 0mm 5mm 10mm, clip=true]{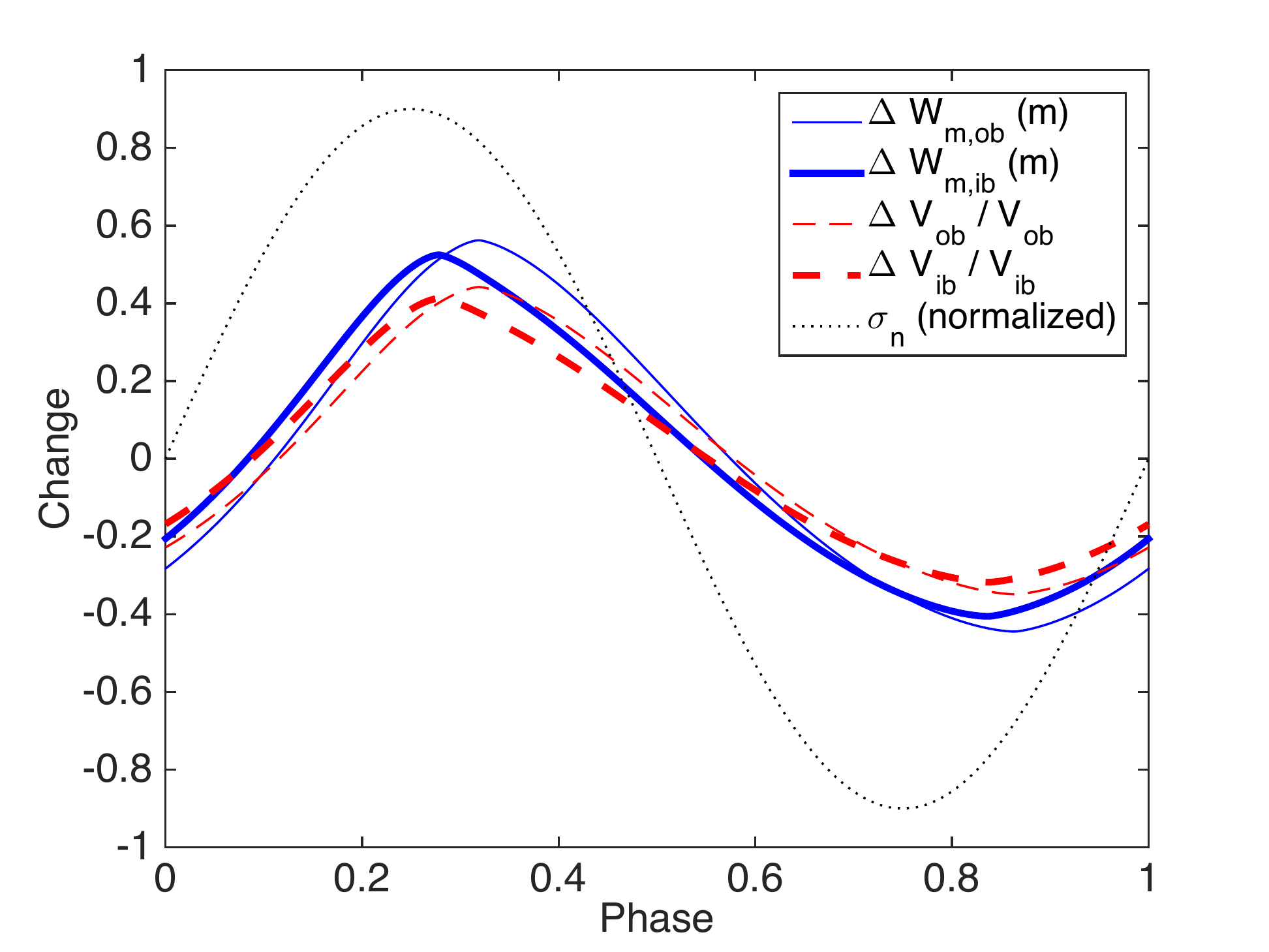}

c)
\includegraphics[width=0.55\textwidth,trim=0mm 0mm 5mm 10mm, clip=true]{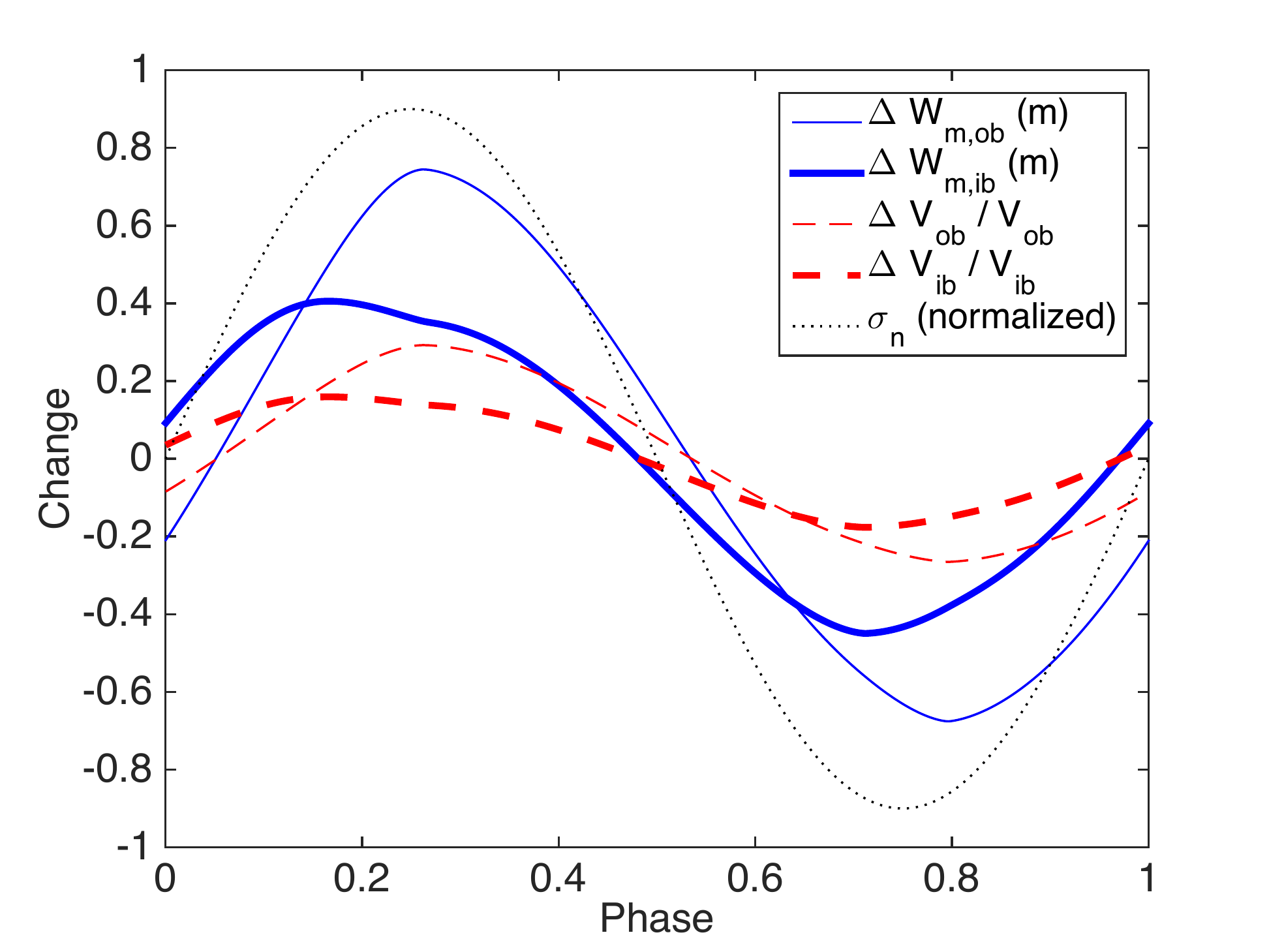}

\end{figure*}

\begin{figure}
\centering
d)
\includegraphics[width=0.65\textwidth]{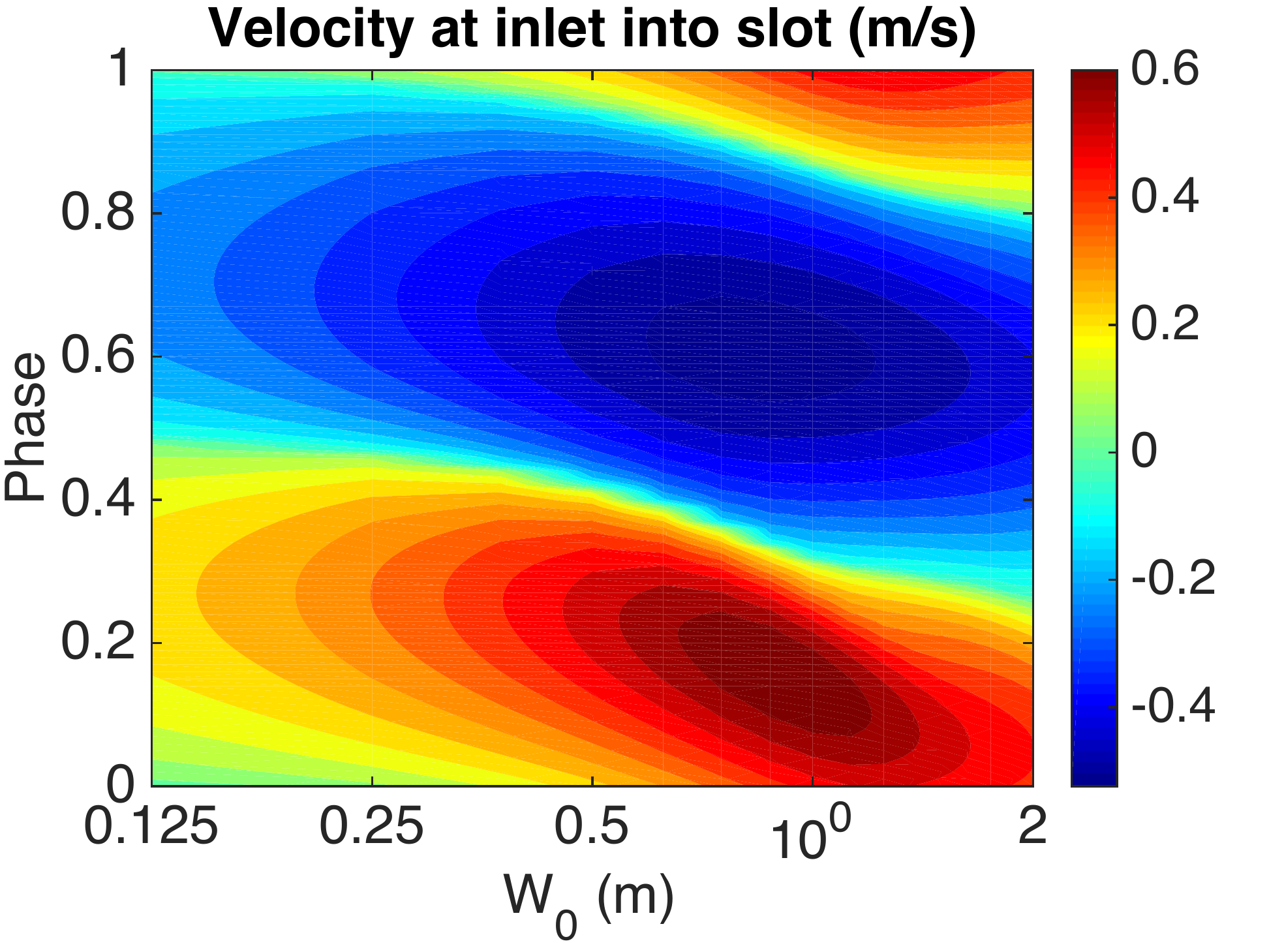}

e)
\includegraphics[width=0.65\textwidth]{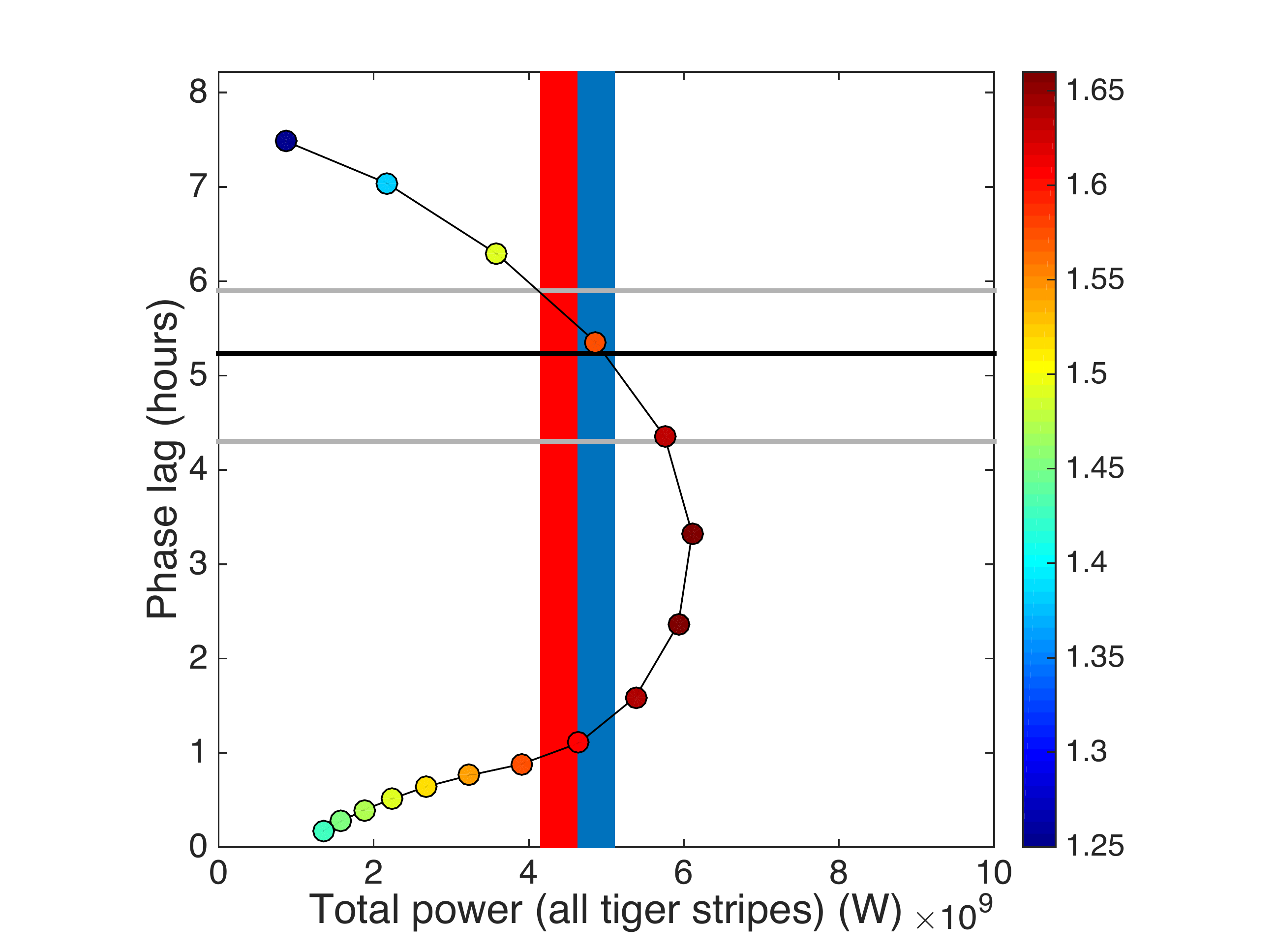}
\caption{
\small
Tidal flexing cycle for interacting slots assuming 2 inboard ($ib$) and 2 outboard ($ob$) slots, $E$ = 6 GPa, $L$ = 100 km. \textbf{(a)} Slot half-width $W_0$ = 0.5 m. $\Delta W_m$ is maximum width change, $\Delta V/V$ is fractional change in slot water volume, and $\sigma_n$ is extensional stress to 90\% of its own peak amplitude. \textbf{(b)} As (a), but $W_0$ = 1.0 m. \textbf{(c)} As (a), but $W_0$ = 2.0 m. Although outboard-slot deformation leads $\sigma_n$, overall power still lags $\sigma_n$ because most power is generated in the inboard slots. \textbf{(d)} Diurnal cycle of velocity into slot at ocean inlet, as a function of initial width. Contour interval 0.05 m/s. Contour interval 0.05 m/s. \textbf{(e)} Overall phase lag versus overall power output. Initial (tidal-stress-free) half-width is sampled at 0.25 m (uppermost dots) and then at 0.125 m intervals up to 2.5~m (lowermost dots). Dot color corresponds to the fractional change in aperture (max/min) during the tidal cycle. Thick black horizontal line shows phase lag relative to a fiducial model, gray lines show 1$\sigma$ error in observed phase lag. Red bar shows the observed power output of Enceladus; blue bar includes the additional power inferred for reheating of cold ice at depth.}
\end{figure}

\subsection{Ice inflow and melt-back}

\noindent In isostatic equilibrium, a differential stress $\sigma_x$ drives viscous ice inflow into the slot (Fig.~S6) \citep{McKenzie2000}. $\sigma_x$ reaches a maximum of $g \rho_i (Z - H)$ at the water table, where $g$~$\approx$~0.1~m/s$^2$ is Enceladus gravity, and tapers linearly to zero at the moon's surface and at the ocean inlet. The corresponding ice-divide strain rate, $\dot{\epsilon}_{xx}(z)$, is given by

\begin{equation}
\dot{\epsilon}_{xx}(z) = \frac{1}{8}  N(T) \sigma_x(z)^3
\end{equation}

\noindent where $N(T)$ is the creep parameter of ice I and the factor of 1/8 assumes confinement in the along-slot ($y$) direction [\cite{CuffeyPatterson2010}, p. 62]. Solid-ice flow is much slower than the oscillating liquid-water flow in the slot. Most of the flow will occur for $T$$~>~$200K, $\sigma_{x}~\sim~10^5$ Pa, and for these conditions $N(T)$ is well-constrained (these flow conditions correspond to terrestrial ice sheets). To solve for $\dot{\epsilon}_{xx}(z)$, we first calculate the ice inflow rate with a conductive geotherm:

\begin{equation}
T(z) = T_{s} + (T_{m} - T_{s})(z/Z)
\end{equation}

\noindent with Enceladus surface temperature $T_{s}$~=~60K and ice-shell base temperature $T_{m}$~=~273K. We log-linearly interpolate $N(T)$ from Table 3.4 in \cite{CuffeyPatterson2010} and approximate the inflow rate $v_x(z)$ as the product of $\dot{\epsilon}_{xx}(z)$ and the half-width between tiger stripes, $S/2$. This gives a peak ice inflow rate of 0.3 m/yr and a depth-averaged ice inflow rate ($\overline{v_x}$) of 0.04 m/yr. However, this is not yet a self-consistent setup. That is because such high ice inflow rates cause rapid subsidence of cold ice from above, which perturbs the geotherm, lowering $T(z)$ \citep{Moore2007}. Meanwhile, the water-filled slot defines an isothermal, relatively warm vertical boundary condition. To take account of these competing effects, we use a 2D conduction-advection code \citep{Gerya2010} to find temperatures within the ice shell  as a function of  a dimensionless P{\'e}clet number $\lambda_z$. 

\begin{equation}
\lambda_z = \frac{\rho_{i} c_{p} v_z Z   }{k_{i}} 
\end{equation}

\noindent where $v_z$ = 2 $\overline{v_x}$ (for $S$ = $Z$) is the subsidence rate, $c_p$ = 2000 J/kg/K is ice heat capacity, and $k_{i}$ = 2.5 W/m/K is ice thermal conductivity. We assume $v_z$ $\neq$ $v_z(z)$, in effect assuming that horizontal flow is concentrated near the base of the slot (this will be justified a posteriori). Then we take the $v_z$ implied by the conductive solution, and use (A16) to get $\lambda_z$, find the corresponding ice-divide temperatures in the 2D model output, and use (A14) to find the new $v_x(z)$. Iteration gives a steady state: $\overline{v_x}$~=~3.4~mm/yr~($Pe_z$~$\sim$~6), peak $v_x$~=~3 cm/yr, melting losses = $2 \overline{v_x} l_{melt} \rho_i D l$~$\approx$~10$^8$~W~per~slot, sensible~heat~losses~=~$2~\overline{v_x}~(273\mathrm{K}~-~60\mathrm{K})~c_p~\rho_i~D~l$~$\approx$~10$^8$~W~per~slot~(Fig. S7). (Here, $l_{melt}$~=~334~kJ/kg is the latent heat of fusion of water). At steady state, with these approximations, 90\% of the inflow comes from the lowermost 20\% of the ice shell, which validates the approximation $v_z$ $\neq$ $v_z(z)$ made above. Faster ice inflow near the base of the shell will narrow the slot at its base, until local enhancement of turbulent dissipation in the liquid-water slot at the narrowed ocean inlet generates enough melt-back to balance inflow.

\begin{figure}
\centering
\noindent\includegraphics[width=0.5\columnwidth]{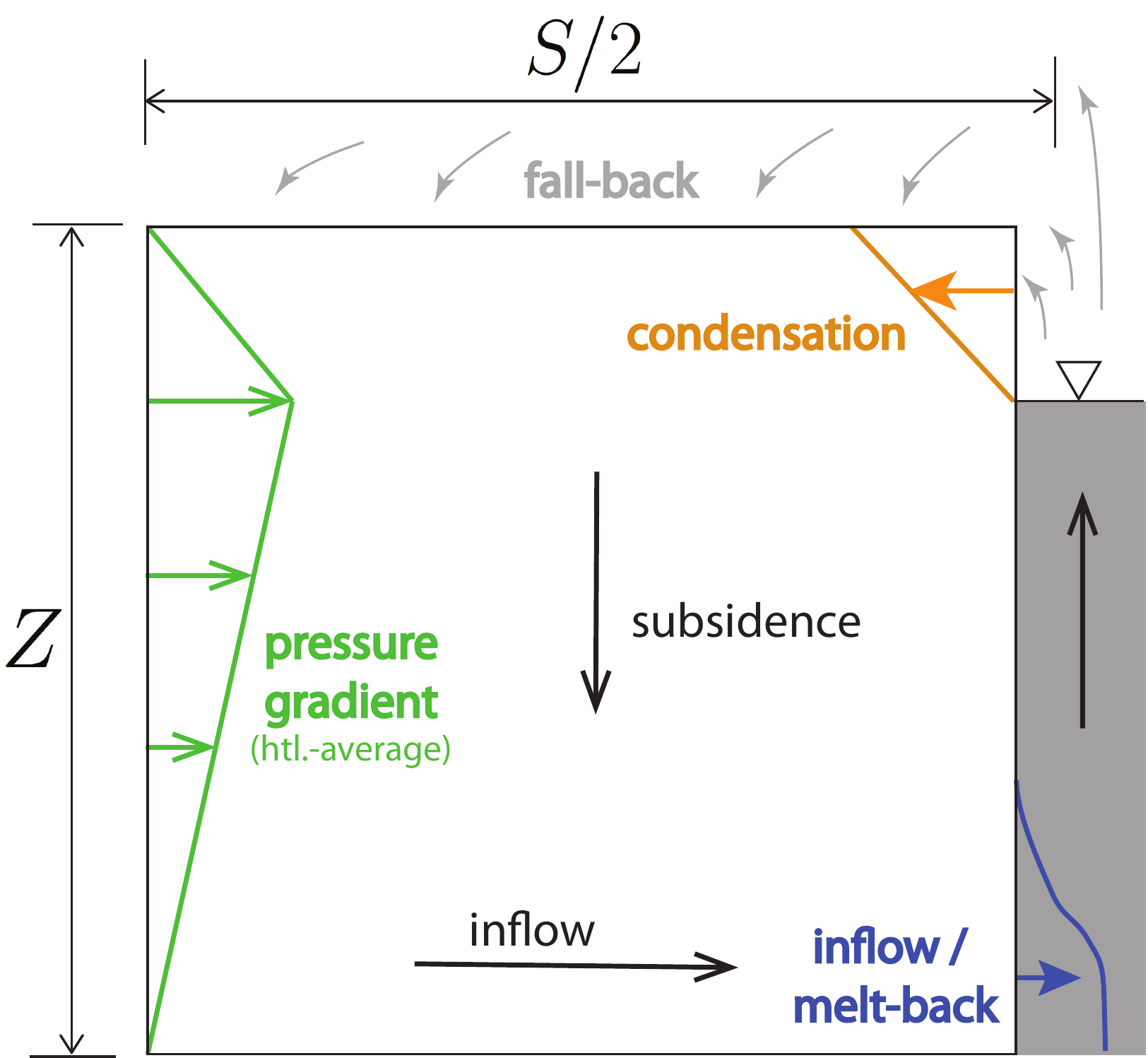}
\caption{Long-lived liquid-water slots (that are in isostatic balance with adjacent ice) set up differential stresses (green arrows show vertical gradient in horizontally-averaged differential stress) that drive flow in the adjacent ice shell. Removal of ice by inflow and slot melt-back is compensated by subsidence. Subsidence provides accommodation space for condensation of vapor and for ballistic fall-back of erupted ice particles (materials that would otherwise seal the slot, powering down the eruptions).}
\end{figure}

\begin{figure}
\centering
\noindent\includegraphics[width=0.7\columnwidth]{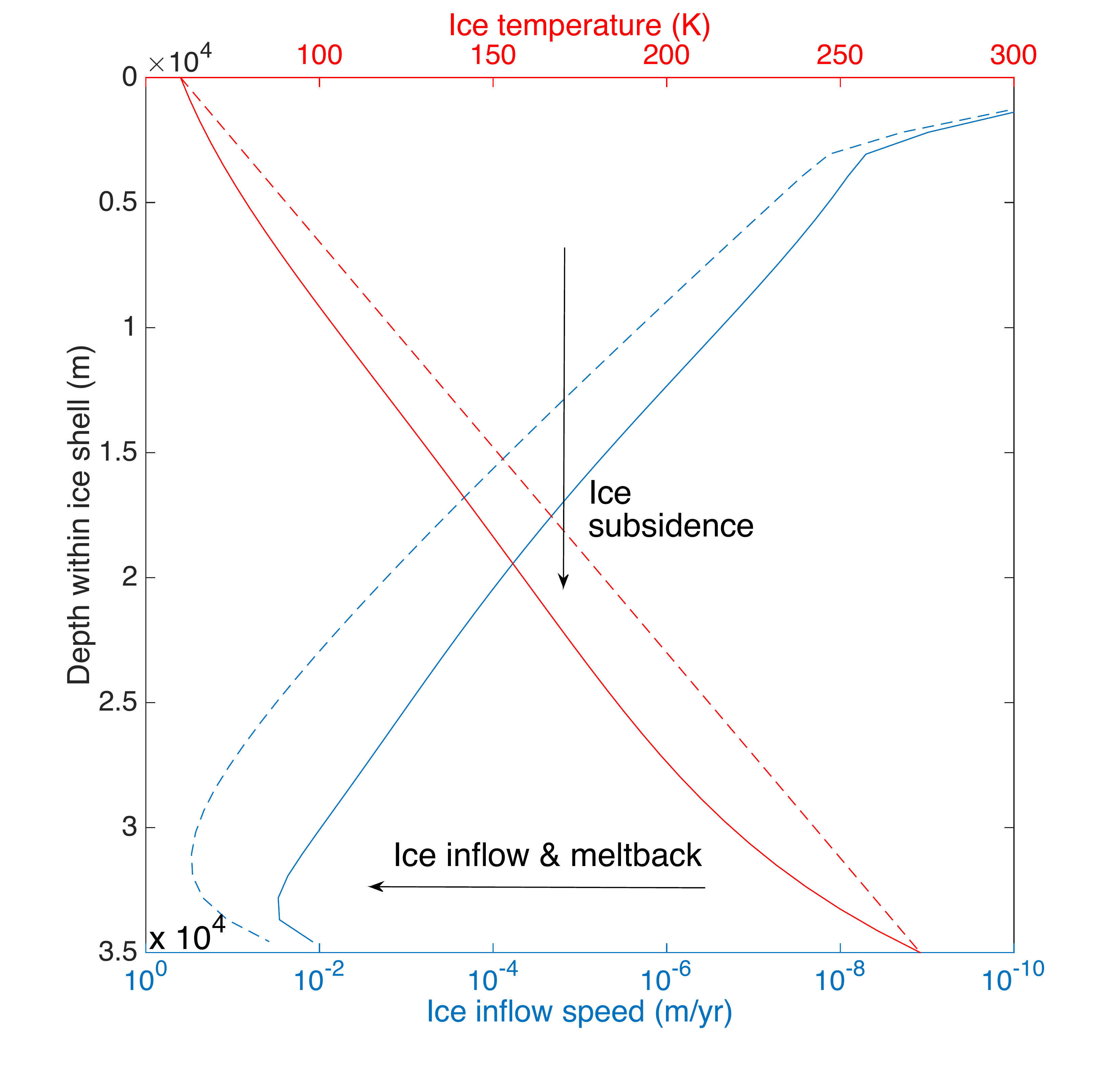}
\caption{Diagram to show how long-lived water-filled slots affect ice shell tectonics. For an initially-conductive thermal profile within the shell (dashed red line), ice will flow viscously into the slot (dashed blue line), driven by the stress difference between the shell and the slot (Fig. S6). The removal of material from near the base of the ice shell causes subsidence of cold ice from above. This cools the lower shell, in turn reducing inflow. Mean ice inflow is reduced ten-fold at equilibrium (solid lines). Shear tractions $\tau_{xz}$ within the ice shell are neglected. Flow is approximated as vertical except in a narrow channel of lateral flow near the base of the shell, which is reasonable because of the large homologous-temperature contrast from the top to the bottom of the shell ($\sim$~0.7) and the strong temperature dependence of the viscosity of ice. 
}
\end{figure}

\begin{figure}
\centering
\includegraphics[width=1.0\columnwidth]{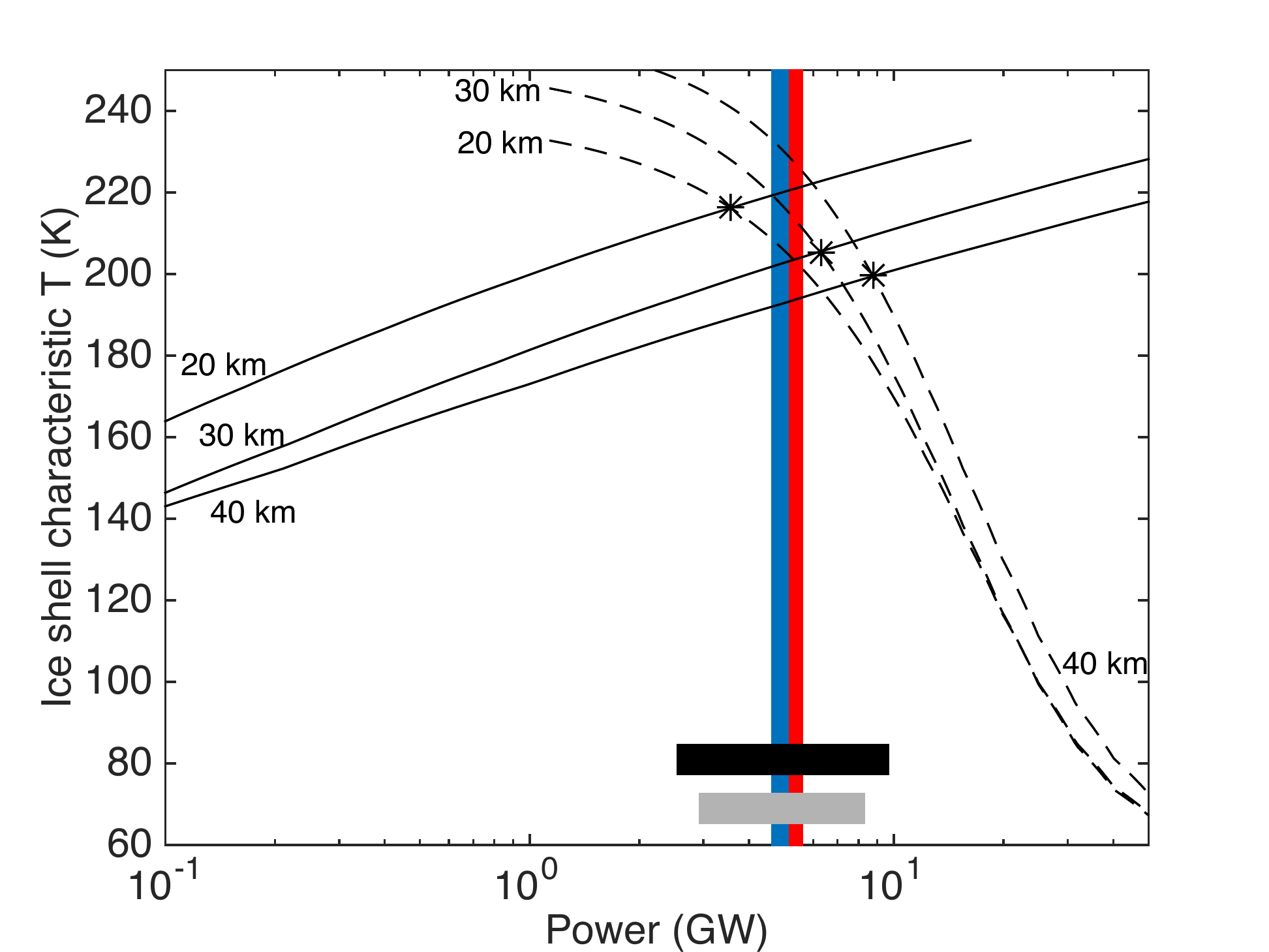}
\caption{\small The power output and mass flux of Enceladus, linking $10^{-2}$~yr through $10^{6}$~yr timescales. Solid black lines show the power corresponding to melting and subsequent vaporization of the ice flowing from the ice shell into base of the slots for ice shell thicknesses of (from top) 20~km, 30~km, and 40~km (the range matching gravity data;\cite{Iess2014}), from our ice flow model. Dashed lines show the cooling of the ice shell corresponding to subsidence of cold ice at that mass flux; where the dashed lines intersect the solid lines, shell equilibrium is possible on Myr timescales (black stars). Line terminations correspond to the ends of our model-output look-up table, not to any physical effect. The ordinate corresponds to ice shell characteristic temperature (the temperature at a depth equal to 80\% of ice shell thickness). The gray bar shows the power output for the turbulent dissipation models that matches the observed diurnal phase lag (relative to a fiducial model) of Enceladus' eruptions to within 2$\sigma$. The black bar shows the range of power output produced by turbulent dissipation model runs that predict a slot aperture that varies by a factor of $>$1.4, consistent with the observation of large-amplitude diurnal variations in plume flux. The million-year and diurnal-average power outputs implied by long-lived slots agree closely both with each other, and with the observed power output of Enceladus (blue bar; red bar includes the additional power inferred for reheating of cold ice at depth).}
\end{figure}

Ice loss by melting near the base of the ice shell is balanced (at steady state) by ice gain near the top of the ice shell by condensation and by frost accumulation. In a horizontal average, subsidence creates accommodation space for the near-surface build-up of condensates (an inevitable consequence of tiger stripe activity, that would otherwise be expected to plug up conduits). The distribution of condensates across the ice surface depends on the poorly-understood mechanics of the uppermost kilometer of Enceladus' ice shell \citep{Davis1983, Helfenstein2008, BarrPreuss2010,  Martens2015}, and on the long-term average partitioning of condensates between ballistic rainout and fissure-wall condensation \citep{Schenk2011}. Because water vapor gives up the latent heat of sublimation $l_{subl}$ upon condensation to ice, we predict a total (four-stripe) tiger stripe terrain thermal emission of $ 4\,\, l_{subl} \,v_z \,\rho_{i} \,L \,S $ $\approx$ 8 GW for $Z$ = 35 km. The~observed~value~is~$\sim$~5~GW including the latent heat represented by the vapor escaping from Enceladus. The short-term observed output is within a factor of $\sim$1.6 of the predicted long-term output for $Z$ = 35 km.

Encouragingly, this roughly correct prediction of the power output of Enceladus requires only $S$, $Z$, $g$, and the material properties of ice. The predicted power output scales as $Z^{\approx 1}$ and intersects the observations at $Z$ = 25 km (Fig. S8). The dependence of power output on $Z$ has the right sign to provide a negative feedback on ice shell thickness.

Ice sheets on Earth deform at rates fit by (A14) with $N(T)$ between 1$\times$ and 5$\times$ values inferred from laboratory experiments (rheological compilation by \cite{CuffeyPatterson2010}). Variations within this range have remarkably little effect on the Enceladus steady-state power output. This buffering suggests that a more sophisticated model of the depth-dependent creep parameter of the ice prism (which might include visco-elasto-plastic rheology, temperature-dependent thermal conductivity, and 3D effects) would not greatly alter the equilibrium values found here.

%

In steady-state geyser tectonics, processes that are not directly observed (melt-back and ice warming) contribute only $(l_{melt}~+~(273\mathrm{K}~-~60\mathrm{K})~c_p)/(l_{vap}~+~l_{melt}~+~(273\mathrm{K}~-~60\mathrm{K})~c_p)$~$\approx$~20\% of the energy demand that is balanced by viscous dissipation. This additional energy corresponds to the offset between the blue and red bars in Figs. 2 and 3. If we are wrong and the tiger stripe terrain has not yet reached tectonic steady state, then the power demand is higher because the bulk inflow rate is greater, but within the envelope of power that can be produced by turbulent dissipation (Fig. 2). \cite{Howett2014} report 4.6$\pm$0.2 GW (since revised to 4.4$\pm$0.2 GW) of excess thermal emission from the tiger stripe terrain - corresponding to a desublimation flux of 1700 kg/s if all of the IR energy is supplied by desublimation. In addition, \cite{Hansen2011} report 200~kg/s of water vapor exiting the moon. Adopting $l_{vap}$ = 2.3~$\times$~10$^6$~J/kg and $l_{melt}$~=~3.3~$\times$~10$^5$~J/g, the directly constrained energy demand is 4.9$\pm$0.2 GW. This corresponds to the red bars in Figs. 2 and 3. In our model, at steady state 2.9~$\times$~10$^5$~J/kg \citep{CuffeyPatterson2010} are needed to re-heat cold ice from surface temperatures ($\sim$60K) to 273K. 

\subsection{Stability of slots.}
\noindent Slot geometry is maintained through evening-out of temperature by along-slot stirring. Stirring is driven by along-slot gradients in the amplitude of diurnal cycles in slot width and flow velocity. These cycles have (for basic slot geometries) peak amplitude near the middle of the slot. Gradients from the middle to the ends of each slot, together with minor along-slot tidal phase variations (Fig. S2), drive bulk along-slot flow with velocity $O$(10\%) that of the vertical flow. 
Because flow within the slot is turbulent, the effective stirring timescale is $O(10\%) L / \overline{V} \sim$ 40 days. Currents in the ocean (which have been calculated at $u$ $>$1 cm/s for Europa; \cite{Soderlund2014}) will sweep water along the base of the slot during the compressive phase of the cycle prior to re-ingestion during the tensile phase of the cycle, and this will help to equalize temperatures within the slot on a timescale $L / u$~($\sim$~100~days for $u$~=~1~cm/s).

These stirring timescales are short compared to the timescales for both melt-back and slot-narrowing via inflow, and this helps to explain why the tiger stripes do not suffer end-freezing nor undergo a corrugation instability. An upper bound on melt-back speed comes from complete dissipation of the tidal stress cycle in the water, for which power per unit volume is $2 A/p$ = 2 W/m$^3$. The corresponding timescale for melt-back doubling of the width of a 1m-wide slot is not less than 1600 days. Halving of slot width by viscous inflow of wall ice also takes much longer than along-slot isothermalization by stirring (\S A.5). This ratio of timescales favors suppression of the fissure-to-pipe transition. This contrasts with magmatic fissures on Earth, for which along-slot thermal homogenization timescales are long compared to freeze-out timescales. As a result, fissure eruptions develop into pipes on Earth. Although a full treatment of instabilities in melt-back will require detailed modeling, this heuristic argument suggests that long-lived fissure eruptions on Enceladus are physically reasonable \citep{Spitale2015}. 

This argument for slot stability conservatively ignores ice-bridge disruption by strike-slip motion ($O$(1 m) per cycle; \cite{SmithKonterPappalardo2008}), which could arrest the conversion of slots to pipes. If pipes do form, they should be short-lived: pipes change their cross-sectional area by a fraction of only ($A$/$E$) $\sim$ 10$^{-5}$ during a tidal cycle, so unless the ocean's pressure undergoes a high-amplitude diurnal pressure cycle, turbulent dissipation of diurnal flow within pipes will be minor and the pipes should quickly freeze. Short-lived pipes may correspond to patches of enhanced emission that have been reported in some analyses of \emph{Cassini} images (e.g. \cite{Porco2014}); along-slot variations in the unfrosted width of fractures about the water table are more consistent with a more recent analysis \citep{Spitale2015}, that does not require pipes.

We have treated the ocean as a constant-pressure bath and ignored feedbacks of flushed water on ocean pressure, which could be significant. 

Fig. 2 in the main text shows how slots can be stable to changes in mean width: on the branch where increases in $W_0$ reduce power output, if dissipation is too low, water will freeze at the margins, narrowing the slot until a steady state can be reached. Similarly, if dissipation is too high, the melt-back rate will increase, widening the slot until a steady state can be reached.

At steady state, the latent heat of vapor escaping the water table must be balanced by heat supplied from the water. Vertical resupply is assisted by exsolving, ascending bubbles \citep{CrawfordStevenson1988}. The km-scale changes in $h$ help: some of the vapor is supplied from ice that is warmed during the time that it is underwater, rather than directly from the water. The changes in water level and in slot width also help to break up any ice that does form. Evaporation makes the water at the top of the slot more salty, favoring subsidence. As slot water evolves to salinity $>$2 wt\%, cooled water would sink even supposing salt content is not increased during evaporation. These processes will buffer water near the top of the slot to modestly elevated salinity.

Observed thermal emission, if conductive, averages over $10^3$ yr timescales because the conductive path length from the fissure walls to the center of the thermal-emission belts is hundreds of meters \citep{Abramov2015}. Thus, the fact that thermal emission from the four tiger stripes is comparable indicates that the spacecraft-era continuity of activity from all four stripes is representative of activity on 10$^3$ yr timescales. The fact that the emission is of the same order of magnitude for the four tiger stripes further suggests a regulating mechanism maintaining the stripes at that power - for example, turbulent dissipation (Fig. 2).

\subsection{Sensitivity tests and extensions.}

\begin{figure}
\noindent\includegraphics[width=0.95\columnwidth,trim=0mm 0mm 0mm 0mm, clip=true]{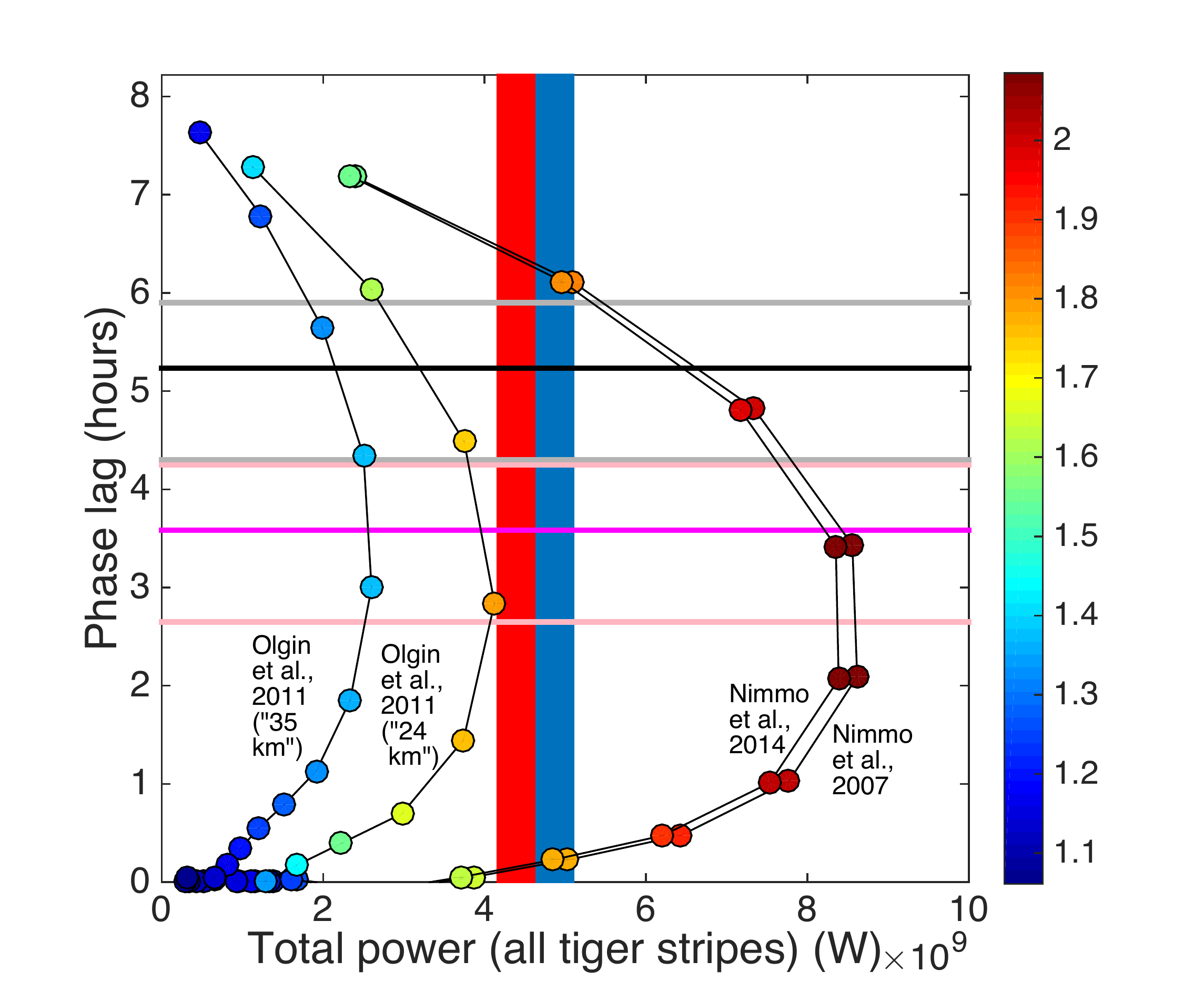}
\caption{\small Sensitivity of time-averaged power to $h_2$ and $l_2$, assuming two outboard slots with $L_{ob}$ = 151 km, and two inboard slots with $L_{ib}$ = 93 km. \cite{Olgin2011} use a four-layer Enceladus structure to convert ice shell thickness to $h_2$ and $l_2$, whereas \cite{Nimmo2007,Nimmo2014} use a three-layer Enceladus structure to convert ice shell thickness to $h_2$ and $l_2$. The numbers in quotes refer to the global ice-shell thicknesses assumed in calculating $h_2$ and $l_2$ (we assume $Z$ = 24 km for the Nimmo et al. models). Thick black horizontal line shows phase lag relative to the \cite{Nimmo2007} model, with gray lines showing 1$\sigma$ error in observed phase lag. Thick magenta horizontal line shows phase lag relative to the \cite{Olgin2011} model for ice shell thickness of 24 km, with pink lines showing 1$\sigma$ error in observed phase lag. Dot color corresponds to the fractional change in aperture (max/min) during the tidal cycle. For each curve, initial (tidal-stress-free) half-width is sampled at 0.25 m (uppermost dots) and then at 0.125 m intervals up to 2.5~m (lowermost dots). Red bar shows the observed power output of Enceladus; blue bar includes the additional power inferred for reheating of cold ice at depth. }
\label{figure_label}
\end{figure}

\begin{figure}
\noindent\includegraphics[width=0.95\columnwidth,trim=0mm 0mm 0mm 0mm, clip=true]{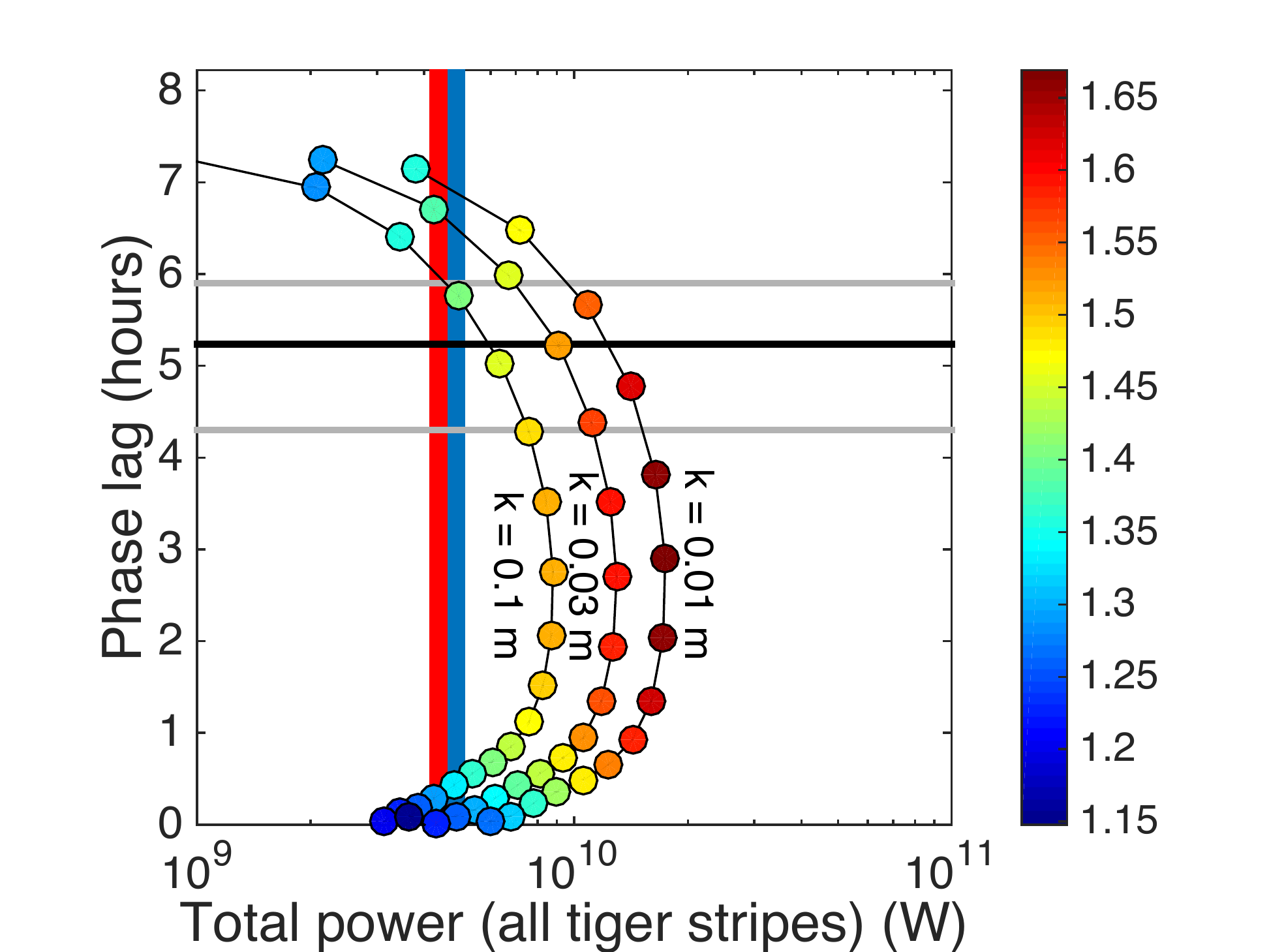}
\caption{Sensitivity of power to $k$ (roughness). Assuming two outboard slots with $L_{ob}$ = 151 km, and two inboard slots with $L_{ib}$ = 93 km. Dot color corresponds to the fractional change in aperture (max/min) during the tidal cycle. For each curve, initial (tidal-stress-free) half-width is sampled at 0.25 m (uppermost dots) and then at 0.125 m intervals up to 2.5~m (lowermost dots). Red bar shows the observed power output of Enceladus; blue bar includes the additional power inferred for reheating of cold ice at depth.} 
\label{figure_label}
\end{figure}

\begin{figure}
\noindent\includegraphics[width=0.95\columnwidth,trim=0mm 0mm 0mm 0mm, clip=true]{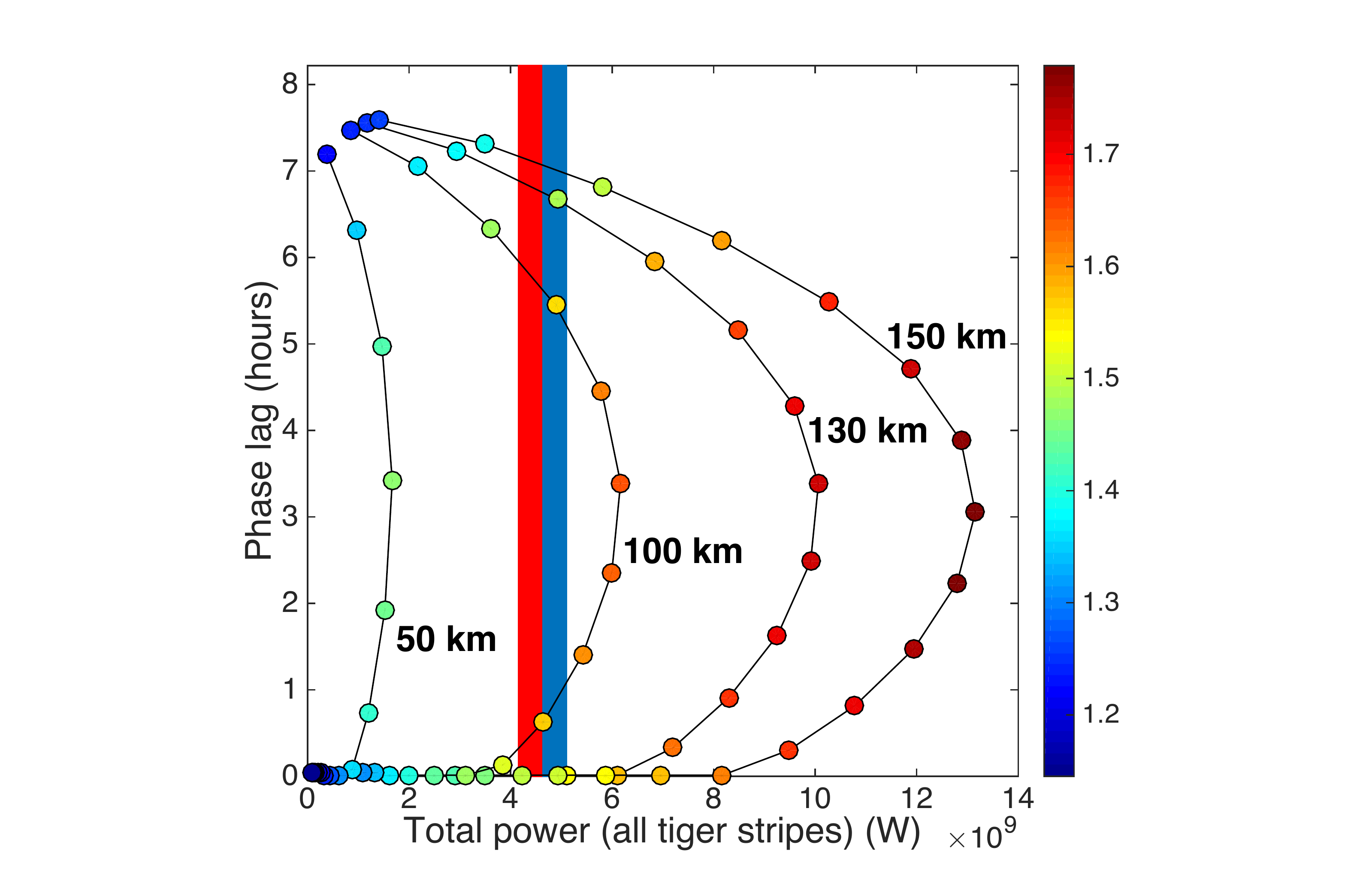}
\caption{Sensitivity of time-averaged power to $L$, considering four non-interacting equal-length slots. Dot color corresponds to the fractional change in aperture (max/min) during the tidal cycle. For each curve, initial (tidal-stress-free) half-width is sampled at 0.25 m (uppermost dots) and then at 0.125 m intervals up to 2.5~m (lowermost dots). Red bar shows the observed power output of Enceladus; blue bar includes the additional power inferred for reheating of cold ice at depth.}
\label{figure_label}
\end{figure}

\noindent$F$ scales as $\sigma_n^2$, and $\sigma_n$ increases with Love numbers $h_2$ and  $l_2$; $h_2$ and  $l_2$ are in turn sensitive to uncertainties in internal structure \citep{Olgin2011,Nimmo2014,Behounkova2015} (Fig. S9). For example, one recent analysis of \emph{Cassini} gravity data gives $z$ $\le$25 km at the tiger stripes \citep{McKinnon2015}. Using the $h_2$ and $l_2$ obtained for a thickness of 24 km by \cite{Olgin2011}, the phase lag is reduced by 1.6 hours relative to the fiducial model employed by \cite{Nimmo2014}. Diurnal stresses and power output are reduced (Fig. S9). Although libration indicates a global ocean and is consistent with high $h_2$ and $l_2$ \citep{Thomas2016}, the remaining uncertainty in Enceladus' true $h_2$ and $l_2$, and the averaging steps in our turbulent power output estimation procedure, mean that agreement between model-predicted peak power and \emph{Cassini} data may be partly coincidental; nevertheless it is encouraging that model predictions bracket the data.


$F$ increases as surface roughness decreases (Fig. S10). Although we are not aware of any $k$ measurements for englacial channels with bi-directional flow, data and modeling of supraglacial channels suggest $k$ $<\sim$ 0.01 m (e.g. \cite{Marston1983}, where we use \cite{AldridgeGarrett1973} to convert between the Manning coefficient and $D_{50}$ and assume $D$ $\sim$ $k$). 

$F$ $\propto$ $L^{\sim 3}$, provided that slots do not interact elastically (Fig. S11). Eruptions appear most concentrated from the central-most 100 km of the tiger stripes \citep{Spitale2015}. Power output for four equal-length 100 km slots is shown in Fig. 2 (left curve).

Nonlinear effects become important for large $L$ for pairs of slots that interact elastically. Here again $F$ $\propto$ $L^{\sim 2}$ is expected, but nonlinear effects become important for large $L$. Consider two slots (or two pairs of slots) $i$ and $j$ whose walls are subject to time-varying internal pressure and which interact through elastic stresses. Neglecting the relatively minor contribution from the $\frac{\partial \Delta W_{m,j}}{\partial t}$ terms, the equations of motion can be written as:

\begin{equation*}
\frac{\partial \Delta W_{m,i}}{\partial t} \sim Q_i + k_1 F(t) + k_2 F(t)(r_{ii}\Delta W_{m,i} - r_{ji} \Delta W_{m,j})
\end{equation*}
\begin{equation} 
\frac{\partial \Delta W_{m,j}}{\partial t} \sim Q_j + k_3 G(t) + k_4 G(t)(r_{jj}\Delta W_{m,j} - r_{ij} \Delta W_{m,i})
\end{equation}

\noindent and the matrix of coupling coefficients is
\[ \left( \begin{array}{ccc}
r_{ii} & -r_{ij} \\
-r_{ji} & r_{jj} \\ \end{array} \right)\] 

\noindent where the $r$ terms are calculated using the output of a boundary-element code (\S A.3). $\left| r_{ij} \right|$ becomes larger relative to $\left| r_{ii} \right|$ as the ratio of tiger stripe spacing to tiger stripe length increases. For slots of the observed length of Enceladus' tiger stripes (Fig. 2), the solution is stable.
For sufficiently long or close-spaced slots the determinant of the coupling matrix becomes negative, and the coupled equations define a saddle-node instability. This instability manifests in the full equations as water piracy. During water piracy, one slot swells, with large-amplitude oscillations, and the other loses almost all its water and undergoes small-amplitude oscillations. 
The pirated slot would eventually become inactive due to reduced turbulent dissipation. Water piracy might limit the size both of Enceladus' tiger stripe terrain and of the three tectonized regions of Miranda, which taken together have angular diameters (70$\pm$10$)^\circ$, $n$ = 4. That is because slots beyond a certain length would suffer destructive interference (if we suppose that the spacing of slots is set by the thickness of the ice shell). Salts on the surface of Europa suggest that conduits link Europa's surface to its sub-ice ocean \citep{BrownHand2013}, and if long water-filled slots on Europa destructively interfered, then this would affect the sustainability of activity \citep{Roth2014}. The importance of tidal flexing to double ridges on Europa has previously been hypothesized by \cite{Greenberg2000}.

\subsection{Links to very long timescales.}
\noindent Our model assumes that tidal heating powers Enceladus, consistent with previous work \citep{NimmoSpencer2013, McKinnon2013, TravisSchubert2015,Hand2011,Vance2007,MalamudPrialnik2013,Porco2006}.

On timescales longer than the tectonic-resurfacing timescale for the South Polar terrain ($<$10$^6$ yr, from crater counts), orbital dynamics require that at least one of the following are true: (1) Enceladus' surface power output is intermittent \citep{MeyerWisdom2007}; (2) the tidal dissipation quality factor of Saturn is $<$1.8 $\times$ 10$^4$, and the mid-sized icy satellites of Saturn are much younger than 4.5 Gyr \citep{Charnoz2010}.

Our work does not address this Gyr-timescale, ``deep'' problem of sustaining ocean-fuelled eruptions on Enceladus. We have focused instead on the ``shallow'' problem of how the eruptions, liquid-water plumbing system, and ice shell tectonics are interrelated at timescales from the 10$^1$ yr observational baseline up to the tiger-stripe terrain's tectonic refresh timescale ($4 L S Z \rho_i l_{subl} / 5\,\mathrm{GW} \approx 10^7$ yr). Beyond the tectonic refresh time geologic information is lost, although terrains far from the tiger stripes \citep{Bland2012} may record Enceladus' activity at earlier times. 


Of the subset of erupted water that goes into orbit around Saturn, very little returns to the tiger stripe terrain. Therefore, if strict steady state in the ice shell is to be maintained, mass escaping from Enceladus should be balanced by mass supplied by the ocean. Tracking changes in ocean volume over millions of years is beyond the scope of this study, but the worst-case imbalance -- supposing that water escaping from Enceladus is entirely sourced from the ice shell, with zero replacement -- is only $\sim$10\% (200~kg/s escapes, 1700~kg/s circulates).

\subsection{Typical velocities within the slot are not much less than the velocity at the slot outlet.}
\noindent Our equations assume the pressure distribution within the slot is linear, so that the flow within the slot is given by a single representative velocity that is not much different from the velocity at the slot outlet. In this section we show that this uniform-pressure-gradient approximation does not affect our conclusion that turbulent dissipation can account for the observed power output of Enceladus.

Flow in the slot is driven by the significant pressure gradient as the water table is displaced from its equilibrium height. In the limit that the water provides negligible resistance to flow (inertial or viscous), the water table remains stationary during the tidal cycle.  Given the tidal stress amplitude of $\sim$100 kPa (peak-to-trough), crack length of $\sim$100 km, and Young's modulus of 6 GPa for ice, the slot width change for no pressure (water table) change within the slot is  $\sim$1.7 m.  From $\partial$$V$/$\partial$z = ($\partial$$W$/$\partial$t)/$W$, with $V$~=~0 at $z$~=~0, $\Delta$$W$$\sim$$W$, $\Delta$$t$ $\sim$ 16 hrs (1/2 tidal cycle), the velocity at the inlet is of order 0.5 m/s, and the average velocity in the slot half that.  In the full solution (equation A6 or A8), the water does provide resistance to flow (this also generates the observed phase lag), such that the water table is displaced between the extreme values that arise from assuming an isolated slot ($\sim$0.5 km for $W$=1 m) and from assuming no resistance to flow (0 m). For our ``best fitting'' results, arising from an equilibrium half-thickness slightly~less~than~1~m, the water table is displaced of order 100 m and the flow velocity at the entrance is of order 0.4 m/s.  

The pressure distribution with height within the slot can be derived as follows. Let $z=0$ be the equilibrium height of water in slot (hydrostatic balance with subsurface ocean). Let  $z=D$ correspond to the base of the ice shell -- the top of the ocean ($z$~positive downward).  Boundary conditions are that the water pressure equals $\rho gD$ at $z=D$ (base of the slot) and zero at $z=h$ (instantaneous height of water). In the following, $P$ is the deviation of water pressure from this hydrostatic case ($P=0$ at $z=D$; $P=\rho gh$ at $z=h$).

For turbulent flow,
\begin{equation}
\label{V}
V=a_1\sqrt{\left|\frac{\partial P}{\partial z}\right|} \ ,
\end{equation}

where the sign of $a_1$ is opposite that of $\partial P/\partial z$.  We assume $\partial W/\partial z$ = 0, although flexure of the ice prism is expected \citep{Reeh1968, Sergienko2010}. From $\partial Q/\partial z=-\partial W/\partial t$
\begin{equation}
\label{Q}
W\frac{\partial V}{\partial z}=-\frac{\partial W}{\partial t} \ ; \ \ \ \frac{\partial V}{\partial z}=-\frac{1}{W}\frac{\partial W}{\partial t}
\end{equation}
\begin{equation}
\label{V2}
V=-a_2z+c_1 \ ; \ \ \ a_2=\frac{1}{W}\frac{\partial W}{\partial t}
\end{equation}
($a_2$ and $c_1$ are functions of $t$).  Merging (\ref{V}) and (\ref{V2}), 
\begin{equation}
\label{rtdPdz}
\sqrt{\left|\frac{\partial P}{\partial z}\right|}=-\frac{a_2}{a_1}z+\frac{c_1}{a_1}
\end{equation}
\begin{equation}
\label{dPdz}
\left|\frac{\partial P}{\partial z}\right|=\left(\frac{a_2}{a_1}\right)^2z^2-2\frac{a_2}{a_1}\frac{c_1}{a_1}z+\left(\frac{c_1}{a_1}\right)^2
\end{equation}

\noindent Our boundary conditions are on $P$, but to integrate (\ref{dPdz}) for $P$ we must know where $\partial P/\partial z$ changes sign.  This occurs at $z=c_1/a_2$.  For convenience, consider the case where $\partial P/\partial z>0$ for $z<c_1/a_2$ (flow near surface is directed upward, in the negative direction; $a_1<0$).  Then for $z<c_1/a_2$
\begin{equation}
\label{P}
P=\left(\frac{a_2}{a_1}\right)^2\frac{z^3}{3}-\frac{a_2}{a_1}\frac{c_1}{a_1}z^2+\left(\frac{c_1}{a_1}\right)^2z+c_2
\end{equation}
Applying the constraint $P=\rho gh$ at $z=h$,
\begin{equation}
\label{c2}
c_2=\rho gh-\left(\frac{a_2}{a_1}\right)^2\frac{h^3}{3}+\frac{a_2}{a_1}\frac{c_1}{a_1}h^2-\left(\frac{c_1}{a_1}\right)^2h \ ,
\end{equation}
so for $z\le c_1/a_2$
\begin{equation}
\label{P2}
P=\left(\frac{a_2}{a_1}\right)^2\left(\frac{z^3-h^3}{3}\right)
-\left(\frac{a_2}{a_1}\right)\left(\frac{c_1}{a_1}\right)\left(z^2-h^2\right)
+\left(\frac{c_1}{a_1}\right)^2\left(z-h\right)+\rho gh \ .
\end{equation}
For $z>c_1/a_2$, multiply the right side of (\ref{dPdz}) by $-1$ and integrate to obtain
\begin{equation}
\label{P3}
P=-\left(\frac{a_2}{a_1}\right)^2\frac{z^3}{3}+\frac{a_2}{a_1}\frac{c_1}{a_1}z^2-\left(\frac{c_1}{a_1}\right)^2z+c_2^\prime \ .
\end{equation}
The constant can be determined by matching the expressions for $P$ in (\ref{P2}) and (\ref{P3}) at $z=c_1/a_2$.

\emph{Case 1}:  For $a_2=0$ ($\partial W/\partial t=0$), and assuming $c_1>0$, $D<c_1/a_2$ and the water is flowing up everywhere.  Applying $P=0$ at $z=D$ yields
\begin{equation}
\label{case1}
\left(\frac{c_1}{a_1}\right)^2=\frac{\rho gh}{D-h} \ .
\end{equation}
$h>0$ means the water table is lowered, consistent with upward flow and $c_1^2>0$.  From (\ref{dPdz}), (\ref{case1}) implies a uniform pressure gradient within the water column.  
For general $a_2$ but still $D<c_1/a_2$, one can again use (\ref{P2}), a quadratic equation for $c_1/a_1$ that provides both the pressure gradient at the inlet from (\ref{dPdz}) and the pressure for elasticity purposes from (\ref{P2}).  

\emph{Case 2:}  For $h=0$, applying $P=0$ at $z=D$ in (\ref{P2}) yields a quadratic equation for $c_1/a_1$ with two imaginary roots.  This is symptomatic of the fact that $\partial P/\partial z$ has become negative for $z<D$, requiring the use of (\ref{P3}).  Matching pressures with (\ref{P2}) at $z=c_1/a_2$ yields
\begin{equation}
\label{c2p}
c_2^\prime=\frac{2}{3}\left(\frac{a_2}{a_1}\right)^2\left(\frac{c_1}{a_2}\right)^3
-2\left(\frac{a_2}{a_1}\right)\left(\frac{c_1}{a_1}\right)\left(\frac{c_1}{a_2}\right)^2
+\left(\frac{c_1}{a_1}\right)^2\left(\frac{c_1}{a_2}\right) + \rho gh \ .
\end{equation}
Inserting this back into (\ref{P3}) and applying $P=0$ at $z=D$ yields a cubic equation for the remaining unknown $c_1$:
\begin{equation}
\label{c2p}
0=-\frac{1}{3}\left(\frac{a_2}{a_1}\right)^2\left[D^3-2\left(\frac{c_1}{a_2}\right)^3\right]
+\left(\frac{a_2}{a_1}\right)\left(\frac{c_1}{a_1}\right)\left[D^2-2\left(\frac{c_1}{a_2}\right)^2\right]
-\left(\frac{c_1}{a_1}\right)^2\left[D-\left(\frac{c_1}{a_2}\right)\right] + \rho gh \ .
\end{equation}

For representative values of $a_1$ and $a_2$ near the peak of the cycle in slot velocity (-0.3882 and 2$\times$10$^{-5}$ s$^{-1}$, respectively), we find that the pressure gradient does diverge from a linear pressure gradient (as anticipated from A23-A28). Pressure gradients are larger near the slot inlet than far from the slot inlet. Because dissipation scales as ($\partial P$/$\partial z$)$^{3/2}$, this implies that the averaged dissipation used in this paper (from A.8 and the linear-pressure-gradient approximation), when compared to the true dissipation, is a mild underestimate.

\end{article}
%
%
%
%
%
%
%
%
%


\end{document}